\documentclass[sigconf]{acmart}

\usepackage{subcaption}
\usepackage{cleveref}
\usepackage{xspace}
\usepackage{tabularx}
\usepackage{booktabs}
\usepackage{multirow}
\usepackage{soul}
\usepackage[font=small]{caption}
\usepackage{graphicx}

\newif\ifdraft
\drafttrue

\newcommand{\modified}[1]{\ifdraft \textcolor{black}{#1}\else\relax\fi}

\newcommand{\CBone}{\emph{Belief Confirmation}\xspace}
\newcommand{\CBtwo}{\emph{Suggester Preference}\xspace}
\newcommand{\CBthree}{\emph{Familiarity Preference}\xspace}
\newcommand{\CBfour}{\emph{Generalization}\xspace}
\newcommand{\CBfive}{\emph{Self-preference}\xspace}
\newcommand{\CBsix}{\emph{Fixation}\xspace}
\newcommand{\CBseven}{\emph{Resort to Default}\xspace}
\newcommand{\CBeight}{\emph{Optimism\slash Pessimism}\xspace}

\newcommand{\CBnine}{\emph{Instant Gratification}\xspace}
\newcommand{\CBten}{\emph{Narrative Preference}\xspace}
\newcommand{\CBeleven}{\emph{Comparison}\xspace}
\newcommand{\CBtwelve}{\emph{Memory bias}\xspace}
\newcommand{\CBthirteen}{\emph{Conspiracy}\xspace}
\newcommand{\CBfourteen}{\emph{Humanization}\xspace}
\newcommand{\CBfifteen}{\emph{Risk Avoidance}\xspace}

\newcommand{\pts}{participants\xspace}

\newcommand{\typeA}{Type \textit{A}\xspace}
\newcommand{\typeB}{Type \textit{B}\xspace}

\newcommand{\new}[1]{\ifdraft\textcolor{black}{#1}\else#1\fi}
\newcommand{\edited}[1]{\ifdraft\textcolor{black}{#1}\else#1\fi}

\AtBeginDocument{%
  }

\setcopyright{acmlicensed}
\copyrightyear{2026}
\acmYear{2026}
\setcopyright{cc}
\setcctype{by}
\acmConference[ICSE '26]{2026 IEEE/ACM 48th International Conference on Software Engineering}{April 12--18, 2026}{Rio de Janeiro, Brazil}
\acmBooktitle{2026 IEEE/ACM 48th International Conference on Software Engineering (ICSE '26), April 12--18, 2026, Rio de Janeiro, Brazil}
\acmPrice{}
\acmDOI{10.1145/3744916.3773104}
\acmISBN{979-8-4007-2025-3/2026/04}

\begin{document}





\title{Cognitive Biases in LLM-Assisted Software Development}

\author{Xinyi Zhou}
\authornote{Both authors contributed equally to this research.}
\affiliation{%
  \institution{University of Southern California}
  \city{Los Angeles}
 \state{California}
  \country{USA}
  }
\email{xzhou141@usc.edu}


\author{Zeinadsadat Saghi}
\authornotemark[1]

\affiliation{%
  \institution{University of Southern California}
  \city{Los Angeles}
   \state{California}
  \country{USA}
}

\email{saghi@usc.edu}

\author{Sadra Sabouri}
\affiliation{%
  \institution{University of Southern California}
  \city{Los Angeles}
   \state{California}
  \country{USA}
}
\email{sabourih@usc.edu}

\author{Rahul Pandita}
\affiliation{%
  \institution{Github Research}
  \city{San Francisco}
   \state{California}
  \country{USA}
}
\email{rahulpandita@github.com}

\author{Mollie McGuire}
\affiliation{%
 \institution{Naval Postgraduate School}
  \city{Monterey}
   \state{California}
  \country{USA}
  }
\email{mrmcguir@nps.edu}

\author{Souti Chattopadhyay}
\affiliation{%
  \institution{University of Southern California}
  \city{Los Angeles}
 \state{California}
  \country{USA}
  }
  \email{schattop@usc.edu}

\renewcommand{\shortauthors}{Zhou, Saghi, et al.}






\begin{abstract}
The widespread adoption of Large Language Models (LLMs) in software development is transforming programming from a solution-generative to a solution-evaluative activity. This shift opens a pathway for new cognitive challenges that amplify existing decision-making biases or create entirely novel ones. One such type of challenge stems from cognitive biases, which are thinking patterns that lead people away from logical reasoning and result in sub-optimal decisions. \emph{How do cognitive biases manifest and impact decision-making in emerging AI-collaborative development?}
This paper presents the first comprehensive study of cognitive biases in LLM-assisted development. We employ a mixed-methods approach, combining observational studies with 14 student and professional developers, followed by surveys with 22 additional developers. We qualitatively compare categories of biases affecting developers against the traditional non-LLM workflows. Our findings suggest that LLM-related actions are more likely to be associated with novel biases.
Through a systematic analysis of 90 cognitive biases specific to developer-LLM interactions, we develop a taxonomy of 15 bias categories validated by cognitive psychologists. We found that 48.8\% of total programmer actions are biased, and developer-LLM interactions account for 56.4\% of these biased actions. We discuss how these bias categories manifest, present tools and practices for developers, and recommendations for LLM tool builders to help mitigate cognitive biases in human-AI programming.

\end{abstract}

%

\maketitle

\section{Introduction}


Large Language Models (LLMs) have fundamentally altered software development. GitHub reports that over 92\% of developers now use AI coding tools~\cite{github_2023_survey}, transforming programming from a primarily generative activity into one dominated by evaluation and curation of AI-generated solutions. This shift represents more than just the adoption of a tool. It constitutes a cognitive revolution in how developers approach problem-solving, make decisions, and construct software systems.

Human decision-making is inherently susceptible to cognitive biases—systematic deviations from rational judgment that can lead to suboptimal outcomes~\cite{tversky1974judgment, kahneman1972subjective}. Cognitive biases are results of systematic inclinations in human thinking and reasoning that often do not comply with the tenets of logic, probability reasoning, and plausibility~\cite{HANSKORTELING2022610, tversky1974judgment, kahneman1972subjective}. In traditional software development, these biases manifest in well-documented ways: confirmation bias leads developers to favor familiar solutions, anchoring bias causes over-reliance on initial implementation approaches, and availability bias skews risk assessment based on recent experiences~\cite{chattopadhyay2020tale}. However, LLM-assisted programming creates a new cognitive landscape where developers must evaluate AI-generated suggestions, formulate effective prompts, and integrate generated code with human reasoning. \emph{We hypothesize that this paradigm shift not only preserves existing biases but amplifies them while introducing novel bias patterns unique to human-AI collaboration.}


As LLMs become integral to software development workflows, understanding their cognitive implications becomes critical for software quality, security, and developer productivity. Biased decisions in LLM-assisted contexts can cascade through entire codebases, introduce subtle bugs that escape traditional testing, and create technical debt that compounds over time. Yet despite the widespread adoption of AI coding tools, we lack a systematic understanding of how cognitive biases operate in these new collaborative contexts.

Prior work has extensively documented cognitive biases in traditional software engineering~\cite{chattopadhyay2020tale, calikli2010empirical} and human-AI collaboration in other domains~\cite{rastogi2022deciding}, but no systematic studies have examined cognitive biases in real-world LLM-assisted programming. Existing bias taxonomies developed for pre-LLM contexts may inadequately capture the cognitive challenges of prompt engineering, solution evaluation, and human-AI collaboration. This gap leaves developers, teams, and organizations without evidence-based strategies for maintaining decision quality in AI-assisted development.

This paper presents the first comprehensive study of cognitive biases in LLM-assisted development. We conducted observational sessions with n=14 student and professional developers, analyzing 2013 development actions including 808 LLM interactions, supplemented by retrospective interviews. To triangulate our findings, we also conducted confirmatory surveys with n=22 developers. 

We began our investigation by adopting the bias categorization proposed in ~\citet{chattopadhyay2020tale} to analyze biases in our dataset. Our analysis revealed a paradox: LLM-related actions are statistically significantly more likely to be biased (53.7\% of all) and more likely to be reversed (29.4\%) based on chi-square tests of independence (with Bonferroni corrections). Yet, traditional bias categories show weak correlation with reversal actions. This suggests fundamental gaps in the traditional bias categories' ability to understand biases when working with LLMs.

To bridge this gap, we conducted a thorough, exhaustive analysis of 239 cognitive biases in collaboration with a cognitive psychologist to identify 90 relevant biases for developer-LLM interactions, grouped into 15 bias categories  (\Cref{sec:rq1}). We then investigate three main research questions: \emph{\textbf{RQ1.} How often do biases appear in LLM-assisted programming? \textbf{RQ2.} What are their impacts? And, \textbf{RQ3.} What strategies can help developers mitigate these biases?}

Our findings revealed that 48.8\% of total actions are biased, and LLM-related actions continue to be statistically more likely to be biased (56.4\%) based on chi-square tests of independence (with Bonferroni corrections). Among these cognitive bias categories, \CBnine and \CBtwo most frequently impact LLM-related actions (19.3\% and 12.6\% respectively). While \CBsix biases led to the largest number of reversed actions (43.4\% of cases), highlighting the need for greater attention to their avoidance. To help overcome biases, we extracted helpful tools and practice suggestions from survey responses for all phases of LLM interactions. Finally, our survey participants indicated that they ``Often'' /  ``Always'' face 12/15 bias categories, with more than 50\% of participants indicating that they find 14/15 bias categories negatively impact their work. We then extracted helpful tools and practice suggestions from survey responses (\Cref{sec:mitigations}).


Our findings echo observations from other recent work on LLM-assisted development, suggesting that the cognitive challenges of software development have not diminished with AI assistance but have instead shifted into new forms requiring updated mental models, practices, and tool designs~\cite{lee2025impact, subramonyam2024bridging}. As the software industry increasingly adopts LLM-assisted workflows, understanding and addressing these cognitive biases becomes essential for maintaining software quality and developer effectiveness in the AI era.

\section{Related Work}

Our research builds on the intersection of two primary areas:
(1) cognitive biases in software development, and
(2) programmer interactions with large language models (LLMs).

\textbf{Cognitive Biases in Software Development and HAI.}
In software development, cognitive factors, including cognitive biases, can lead to suboptimal practices and outcomes~\cite{11029967}. Studies by \citet{mohanani2018cognitive}, \citet{calikli2010empirical}, and \citet{chattopadhyay2020tale} have examined how biases \modified{as} anchoring and confirmation bias affect developers' estimations, decision-making, and real-world programming scenarios, highlighting the importance of recognizing and mitigating these biases.
In Human-AI Interaction, biases have been studied in explainable AI~\cite{ha2024improving}, AI-assisted decision-making~\cite{rastogi2022deciding}, and voice assistants, typically focusing on common biases \modified{as} confirmation and automation bias. Coding assistant studies~\cite{barke2023grounded} examined behavior patterns with only a brief mention of anchoring bias. Our work extends \citet{chattopadhyay2020tale} by categorizing a broader range of cognitive biases specifically in LLM-assisted programming contexts, examining their impacts on developer-LLM interactions.

\textbf{Programmer-LLM Interaction.}
The integration of large language models into programming workflows has transformed how developers approach coding tasks.
Research by \citet{ross2023programmer} explored conversational interactions between programmers and LLMs, a study done by \citet{nam2024usinganllm} revealed a bright future of in-IDE prompt-less interaction between programmers and LLMs, and other studies demonstrated significant productivity gains when developers collaborated with LLMs, highlighting the efficiency and effectiveness of AI-assisted programming~\cite{jin2024llms}.
However, some studies found challenges in this interaction \cite{sabouri2025trust,sallou2024breaking}.
\citet{sabouri2025trust} found an over-trust that happens due to optimism toward LLM for software developers, which resulted in unwanted actions. 


%

\section{Methodology}

\begin{figure*}
    \centering
    \includegraphics[width=\linewidth]{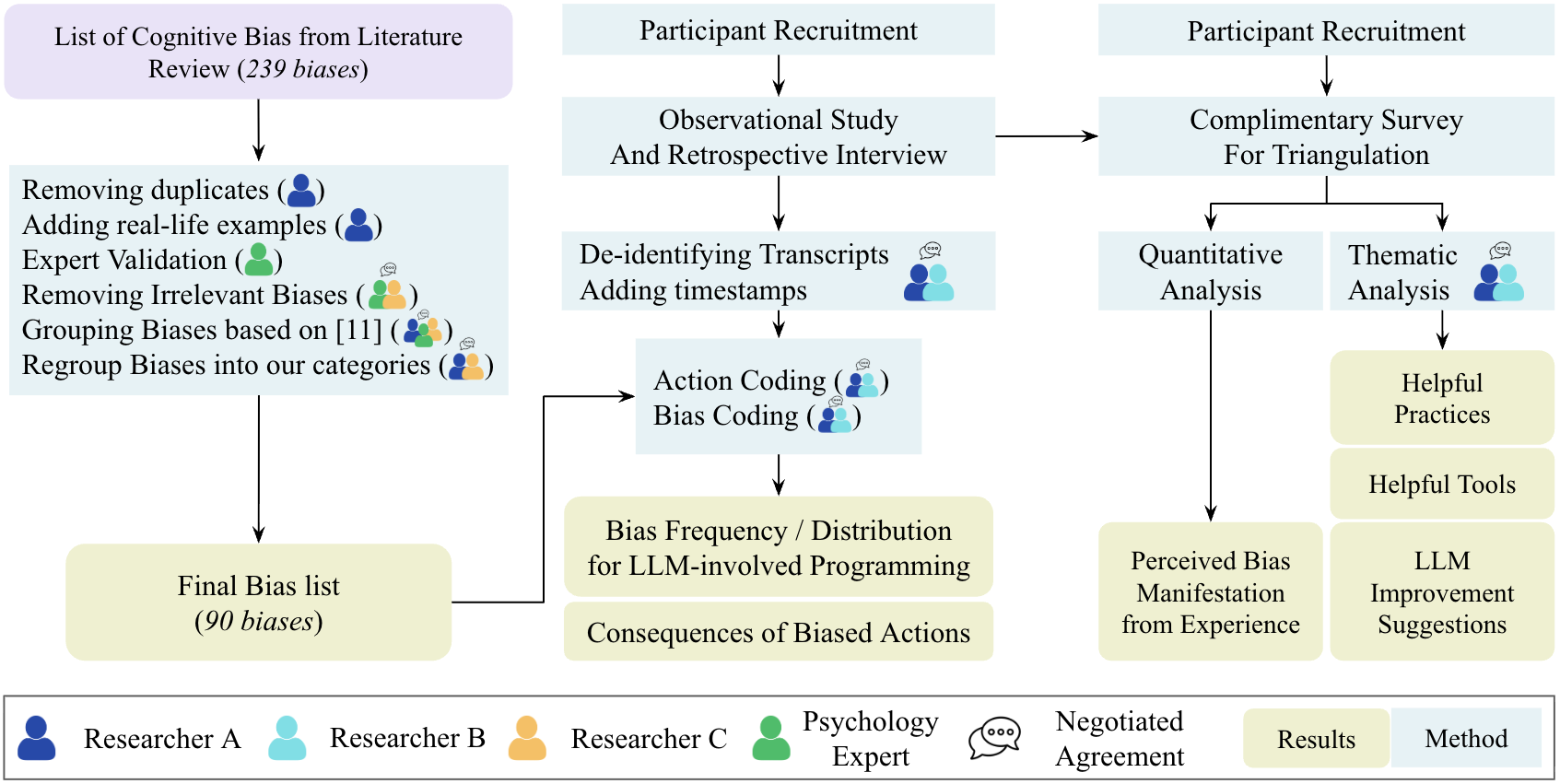}
    \caption{Methodology Overview Diagram}
    \label{fig:overview}
    \Description{Overall methodology diagram}
\end{figure*}

Our methodology contains three components---the approach we used to construct our cognitive bias categorizations, the procedure of observational study sessions, and the survey process used to explore mitigation strategies and \modified{verify} our findings. \modified{All studies were approved by USC Institutional Review Board (IRB).} Figure~\ref{fig:overview} shows an overview diagram of our methodology steps.

\subsection{Bias Category Extraction}\label{sec:taxonomy}

We analyzed our data based on two different bias categorizations.

\paragraph{Bias Categories from non-LLM development workflow.}
To analyze how cognitive biases manifest in LLM-assisted programming, we adopted a bias categorization relevant to software development prior to LLM introduction. We draw on the framework proposed by Chattopadhyay et al.~\cite{chattopadhyay2020tale}, which identified 10 cognitive bias categories in non-LLM programming contexts. Adopting their final categorization enabled consistent comparison between traditional and LLM-assisted programming environments. Throughout this paper, we refer to these categories as $preCB_1$ to $preCB_{10}$. 

\paragraph{Extracting expanded bias categories for LLM-assisted development.}
While prior work~\cite{chattopadhyay2020tale,mohanani2018cognitive} reported traditional software development related biases, it does not account for new biases that may emerge in human-AI collaborative development. For instance, Group Attribution Bias---the tendency to generalize traits or behaviors of an entire group based on impressions of a single individual--has been observed in human interactions with LLM agents~\cite{Echterhoff2024CognitiveBI}. This highlights the importance of identifying additional biases that may arise in the developer’s decision-making process when working with LLMs.
\modified{To capture these biases, we combined the 37 biases from the systematic mapping study of~\citet{mohanani2018cognitive} with Wikipedia's ``List of Cognitive Biases\footnote{\url{https://en.wikipedia.org/wiki/List\_of\_cognitive\_biases}
accessed February 17, 2024}'' to identify 239 biases. A review of 40 recent papers (e.g., software engineering~\cite{barke2023grounded}, psychology~\cite{baron2023thinking}, explainable AI~\cite{ha2024improving}) found no additional biases. While no definitive index exists, over 200 biases have been documented~\cite{mohanani2018cognitive}, and a cognitive psychologist confirmed no biases were missing.}
 
We analyzed biases based on their impact on developers programming with AI assistants. Since multiple biases can have similar effects, we group them by how they influence decision-making, following the method outlined in prior work~\cite{chattopadhyay2020tale, mohanani2018cognitive}.

\emph{Process.} After removing eight duplicate biases such as Hindsight Bias, \modified{researcher A} used bias form references, dropped one bias with contradictory definitions, and constructed examples of how each bias manifests during software development. After \modified{researcher A} reached a negotiated agreement with a cognitive psychologist \modified{with eight years experience}, they excluded 77 irrelevant biases \modified{such as} Pareidolia--which occur in image recognition \cite{voss2012potato}--and human collaboration biases that don't apply to human-AI collaboration~\cite{Felin2024TheoryIA}. The \modified{psychologist} and \modified{researcher B} then annotated and voted to reach a negotiated agreement to tag 90 of the remaining 153 biases as cognitive biases and 63 as heuristics, logical fallacies, or cognitive effects. \modified{Researcher A, B, and psychologist individually categorized the 90 biases. Through iterative inductive coding and negotiated agreements~\cite{garrison2006revisiting}, 15 final bias categories were reached.} The categorization is presented in \Cref{bias-categorizations} and explained in \Cref{sec:rq1}. Definitions of the 239 biases are in the supplementary materials\footnote{Supplementary Materials: \url{https://figshare.com/s/db8f934f65c28ab3eac5}} along with a comparison of our categories with prior work.

\subsection{Observational Studies}
To examine the impact of the bias categories, we conducted hour-long observational sessions on Zoom meetings with 14 developers\footnote{Zoom has been used as a tool for a similar observation study in the ICSE~\cite{Fronza2022KeepingFA} research papers.}.


\emph{Recruitment and Screening.}
We recruited student developers from \modified{one} university through IRB-approved emails and professional developers via snowball and convenience sampling from industry contacts in various organizations. Participants were required to have at least 3 years of programming experience (for students) or professional experience (for developers), and use LLMs for programming tasks, which is crucial for our study. We conducted an initial screening survey (provided in supplementary materials), receiving responses from 153 participants (144 students + 9 professionals). We excluded participants with under 3 years of experience and selected a balanced sample by LLM usage level (``occasionally,'' ``regularly,'' ``frequently''). \modified{We recruited 14 screened participants. We excluded candidates with $<3$ years experience and balanced choices by LLM usage level. After behavioral pattern saturation was reached with seven professionals (P8-14), seven students (P1-7) were randomly selected to balance the participant groups.} \Cref{Demographic} shows demographic information about the participants, including programming experience, preferred languages, and LLM assistants used. The sample is diverse in terms of programming languages, industry (e.g., medical, \modified{life science}, software), organization size (small startups to multinational firms), and roles.



\emph{Protocol.}
Participants brought personal or open-source projects into the sessions to ensure naturalistic LLM interactions and bias observation. \modified{Sessions were conducted individually where each participant was assigned a zoom slot with one of the researchers for observation. }Each session comprised two parts: a 60-minute observational period where participants coded while thinking aloud (following \cite{chattopadhyay2020tale}), with researchers taking notes on notable actions, followed by a 15-20-minute retrospective interview with clarifying questions about their intentions, behaviors, and past LLM coding experiences. All sessions were \modified{audio and screen }recorded with participant consent for subsequent analysis.

\begin{table}
    \scriptsize

  \vspace{-1em}
\def\arraystretch{1} 
    \caption{Demographics of Field Study Participants}
  \resizebox{\linewidth}{!}{
  \begin{tabular}{llllp{1.5cm}ll}
    \toprule
    P\#&Gnd. & Exp./Prof& Prog. Lang(s).& LLM & LLM Freq.& Bg\\
    \midrule
    P1&F&3-6 yrs&HTML, CSS, JS & ChatGPT & Occasionally & Student\\
    P2&M&3-6 yrs&HTML, CSS, JS & GitHub Copilot & Regularly & Student\\
    P3&M&$>$6 yrs &HTML, CSS, JS & GitHub Copilot, ChatGPT, Gemini & Frequently & Student\\
    P4&M&$>$6 yrs &Python & ChatGPT & Frequently & Student\\
    P5&F&3-6 yrs &Python & ChatGPT, Claude & Frequently & Student\\
    P6&M&$>$6 yrs &Rust & GitHub Copilot & Frequently & Student\\
    P7&M&3-6 yrs &Swift & ChatGPT & Regularly & Student\\
    
    P8&M&3-6 yrs&Python & ChatGPT & Occasionally & L Software\\
    P9&M&3-6 yrs&HTML, CSS, JS & Microsoft Copilot & Regularly & L Software\\
    P10&M&3-6 yrs&HTML, CSS, JS & ChatGPT & Occasionally & M Life Science\\
    P11&M&3-6 yrs&HTML, CSS, JS & ChatGPT & Regularly & M Life Science\\
    P12&F&$>$6 yrs&Python & ChatGPT & Occasionally & L Software\\
    P13&M&$>$6 yrs&Python & Copilot, Phind & Rarely & S Medical\\
    P14&M&$>$6 yrs&SQL & ChatGPT & Rarely & M Life Science\\
    \bottomrule
  \end{tabular}}

  \parbox{0.99\linewidth}{%
        \small
        \centering
        \modified{P\#: Participant, Gnd: Gender, Exp: Years of experience, Prog. Lang(s): Programming Languages, LLM: familiar LLM, LLM Freq: LLM usage frequency, Bg: background (S : small, M: medium, L: large company size).}
        }

    \label{Demographic}
\end{table}

\subsection{Qualitative Analysis}

\modified{Two researchers (A, C)} transcribed and classified actions and biases from each session. We recorded timestamped data including: (1) participant actions (e.g., ``removed padding from email text field in \texttt{navigation.css}''), (2) LLM prompts and generated code descriptions, and (3) participant verbalization. Each timestamp was coded for action description, verbalization, reversal status, generated code, LLM interaction mode, and associated bias category.

We define an \emph{action} as a distinguishable step taken by a participant toward a single, consistent goal or intention. Actions are established units for studying developer behavior, including productivity~\cite{meyer2014software}, coding patterns~\cite{minelli2015know}, and cognitive biases~\cite{chattopadhyay2020tale}. While Chattopadhyay et al. identified five traditional programming actions (\textit{Read}, \textit{Edit}, \textit{Navigate}, \textit{Execute}, \textit{Ideate}),  programming now includes diverse LLM interactions through chat interfaces, code comments, and other modalities.
To avoid single-researcher bias, three authors collaboratively coded all participants' actions. We randomly selected 30 minutes from all sessions and iteratively developed distinct action categories through negotiated agreement. The final taxonomy comprises nine action codes (\Cref{action-coding}), extending categories to capture LLM-specific interactions.


We adopted \emph{reversal action} from prior work~\cite{chattopadhyay2020tale}, defining reversal actions as ``actions that developers need to undo, redo, or discard at a later time,'' indicating ``non-optimal solution paths.'' This metric quantifies the impact of biases on developer time and effort by identifying decisions that lead to backtracking or discarded work.
\begin{table}
  \scriptsize

  \caption{\modified{Action Codes and their Definitions}}\label{action-coding}
  \vspace{-1em}
  \label{tab:commands}
  \def\arraystretch{1} 
  \resizebox{\linewidth}{!}{
  \begin{tabular}{p{1.7cm}p{5cm}}
    \toprule
    Action&Definition\\
    \midrule
    Read& Examining information from artifacts (e.g., code, documentation, LLM output).\\
    Edit& Changing the code directly without using LLM\\
    Navigate& Moving within or among artifacts (e.g., online browsing, opening files, scrolling through a file).\\
    Execute&Compiling and/or running code/command.\\
    Ideate&Constructing mental model of future changes.\\
    Inquire& Writing prompts for LLM tools\\
    Adopt Suggestion&Following LLM suggested instructions(e.g., copy and paste codes from LLM)\\
    Ignore Suggestion&Disregarding suggestion instructions or codes generated by LLM tools\\
    Edit w/ Autocompletions& \modified{Changing} the code directly with help from inline or chat-based LLM\\
    \bottomrule
  \end{tabular}}
\end{table}

\modified{Researcher A} and an \modified{NLP} expert \modified{with 3 years experience} collaboratively mapped out the space for \emph{developer-LLM interaction modes} based on the interaction interface and input/prompt content. They identified seven interaction modes such as \verb|a\n| LLM autocompletes code after the developer enters a new line, predicting subsequent lines, \verb|a\code| LLM autocompletes after the developer writes some code manually, and \verb|c\img| when developers include images in the prompt, and \verb|c\img + c\code| when developers include both image and code in the prompt (See supplementary for details).


We examined all participants' actions, verbalization, and LLM interactions to identify and label associated old and new \emph{bias categories} (categories are discussed in \Cref{sec:rq1}). Two authors iteratively coded the bias categories, discussing and updating the categories until they reached an agreement~\cite{garrison2006revisiting}.

\begin{table}[htp!]
  \caption{Demographics of Survey Participants }
  \label{Survey-Demographic}
\def\arraystretch{1} 
  \resizebox{\linewidth}{!}{
  \begin{tabular}{llll|llll}
    \toprule
    SP\# & Gnd. & PE. & Prog. Lang(s). & SP\# & Gnd. & PE. & Prog. Lang(s). \\
    \hline
    SP1  & M  & $<$3 & JS                      & SP12 & F  & $>$6 & C, C++, etc. \\
    SP2  & M  & 3-6  & JS                     & SP13 & M  & $>$6 & JS \\
    SP3  & M  & $>$6 & Python, C\#            & SP14 & M  & $<$3 & HTML, Python, etc. \\
    SP4  & M  & $>$6 & Kotlin, Java, etc.      & SP15 & M  & $<$3 & Python, JS, etc. \\
    SP5  & F  & 3-6  & Java                   & SP16 & M  & 3-6  & Typescript, PHP \\
    SP6  & M  & $<$3 & Python, JS, etc.        & SP17 & M  & $>$6 & SQL \\
    SP7  & M  & $>$6 & Python, JS              & SP18 & F  & $<$3 & Python \\
    SP8  & F  & 3-6  & Python, C++, etc.      & SP19 & M  & $<$3 & Python, JS, etc. \\
    SP9  & M  & 3-6  & Python, Ruby, etc.     & SP20 & M  & $<$3 & SQL, R \\
    SP10 & M  & 3-6  & C++, C\#, etc.         & SP21 & M  & 3-6  & PHP \\
    SP11 & M  & 3-6  & JS, Python, etc.       & SP22 & M  & $<$3 & Python \\
    \bottomrule
  \end{tabular}}

  \parbox{0.99\linewidth}{%
        \small
        \centering
        \modified{SP\#: Survey Participant, Gnd: Gender, PE.: Years of professional experience, Prog. Lang(s): Programming Languages}
        }
\end{table}

Two researchers \modified{(A, C)} transcribed and coded the post-session interviews. \modified{They} unitized the interviews at a conceptual level and collaboratively annotated the participants' \emph{attitudes} based on the LLM usage patterns.

\subsection{Complementary Survey}

We surveyed 22 developers to triangulate observational findings and explore bias mitigation strategies. \modified{Developers were selected because their firsthand insight gives us practical, context-aware mitigation solutions.}
We conducted surveys instead of interviews to gather insights from a larger population while minimizing researcher presence effects (which leads to the Hawthorne effect). 
Participants were recruited on Prolific using screening criteria targeting employed coding professionals with prior AI tool experience. \modified{Of 175 qualified candidates, 24 responded; 22 were retained after excluding 2 low-quality responses.} All \modified{participants} regularly used LLM tools (sometimes: n=4, often: n=11, always: n=7) and primarily worked as Software Engineers/Developers (63.63\%) or Data Analysts (22.7\%), spanning varied experience levels (\Cref{Survey-Demographic}). The survey included 15 scenario-based questions covering different bias categories, and \modified{two researchers (A, B) drafted the scenarios using definitions and examples in Table~\ref{bias-categorizations}.}


\modified{We revised ambiguous wording and ensured each scenario was phrased neutrally, without positive or negative cues. We conducted five expert pilot surveys to iteratively refine the scenarios until participant interpretations aligned with our intended descriptions.}
\modified{One example of these scenarios:} Charlie rejecting an LLM suggestion without thorough evaluation. For each scenario, respondents assessed: ``\emph{How often do software developers experience such scenarios?}'' and ``\emph{How do such scenarios impact the software development process?}'' We also collected mitigation suggestions for tools, LLMs, and practices.

Frequency and impact responses were analyzed quantitatively, while mitigation suggestions underwent Pattern Coding by two researchers \modified{(A, C)} with negotiated agreement~\cite{mathew1994qualitative}. \modified{During the Pattern Coding, they first individually coded all answers, then negotiated and reached agreement on the first-level coding. For disagreements, they introduced a third researcher to vote. In the second round of coding, first-level codes were merged and hierarchical themes emerged for each CB. }Two analysis rounds identified four key categories of helpful practices and tool/LLM improvements for bias mitigation (\Cref{sec:mitigations-tools}).

\section{Overview of Participants' Session}
We observed 2,013 actions across all sessions, with an average of 144 actions per participant and 58 actions in interaction with an LLM. During the observation session, participants were given the option to use an LLM of their choice. They selected a range of LLMs---ChatGPT (10$\times$), GitHub Copilot (5$\times$), Microsoft Copilot (1$\times$), Phind (1$\times$), Gemini (1$\times$), and Claude (1$\times$)). Four participants [P3,5,9,13] used more than one LLM. The dominant interaction type was \verb|c\text| (chat-based), used by all 14 participants, with a range of 7 (P11) to 32 (P3). 

\modified{We analyzed participants’ actions with and without LLMs. The most common action was \textit{Edit} (351 instances) and the least common was \textit{Ignoresug} (14). For LLM-related actions, \textit{Inquire} was most frequent (228). In coding actions, \textit{Navigate} followed Execute (266). \textit{Read} was also common across both categories. Participants worked on varied tasks, including Unity game development [P1], stock tracking [P2], UI development [P6,8–10], graph analysis [P4], data manipulation [P12], and SQL schema design [P14].}




\section{Biases in programming with LLMs}


To understand how cognitive biases impact programmers when working with LLMs, we first adopted Chattopadhyay et al.'s~\cite{chattopadhyay2020tale} bias categorization. 
\modified{Two researchers ($\text{B, C}$) independently coded each action using the 10 bias categories ($\text{preCB}_1-\text{preCB}_{10}$). They then discussed and resolved all coding conflicts by referring to definitions and participant data (quotations/interviews) until negotiated agreement was reached.}
We identified eight of the 10 $preCB$ categories in our 2013 actions: [$preCB_1$] selecting actions based on preconceptions, [$preCB_2$] preferring self-crafted artifacts, [$preCB_3$] fixating on initial information, [$preCB_5$] being optimistic without caution, [$preCB_6$] selecting simpler paths, [$preCB_7$] relying on tools without evaluation, [$preCB_9$] choosing actions based on appearance, and [$preCB_{10}$] affecting developer information recall. Table~\ref{tab:old_CB_count} presents statistics including action counts and affected participants. We did not observe [$preCB_4$] (choosing default options) and [$preCB_8$] (ignoring information).

\begin{table}[ht]
\centering
\caption{ \modified{Frequency and Participant Distribution}}
\vspace{-1em}
\footnotesize
\begin{tabular}{|l|l|c|l|}
\hline
\textbf{Bias} & \textbf{Name} & \textbf{Count} & \textbf{Participants} \\
\hline
$preCB_1$  &  Preconceptions      & 152 & 1--2, 4--5, 8--13 \\
$preCB_2$  &  Ownership      & 139 & 3, 6--8, 11, 13--14 \\
$preCB_3$  &    Fixation     & 298 & 1--5, 8--9, 12, 14 \\
$preCB_5$  &    Optimism    & 96  & 1, 3--5, 9--14 \\
$preCB_6$& Convenience & 290 & 1--11, 13--14 \\
$preCB_7$& Subconscious Action     & 338 & 1--5, 7--13 \\
$preCB_9$& Superficial Selection   & 10  & 1--2 \\
$preCB_{10}$& Memory Bias             & 177 & 1, 4, 6--11, 13--14 \\
\hline
\end{tabular}
\label{tab:old_CB_count}
\end{table}








\subsection{Comparing actions with and without LLMs}
We also replicated the analysis in the prior work~\cite{chattopadhyay2020tale} to understand what proportion of biased actions lead to negative outcomes. We adopted their notion of \emph{Reversal Actions}, the actions that developers need to undo, redo, or discard at a later time. Figure~\ref{fig:pre-bias-rev-llm} (a) presents the distribution of biases and reversal actions in our dataset. Of the 2,013 actions observed, 949 actions could be associated with at least one $preCB$ category, which amounts to 47.14\% of actions. This is comparable to \citet{chattopadhyay2020tale}'s finding that 45.72\% of 2084 actions are associated with at least one category. However, the number of reversal actions in our dataset is only 25\% (505/2013) actions compared to almost a 52.9\% reversal action rate (1104/2084 actions) reported in the previous study. Following the prior work's analysis, we further conducted a Chi-square test of independence with a Bonferroni correction (to account for multiple comparisons) to understand the bias-reversal association. The chi-square test revealed no significant association between biased actions and reversal actions ($p-value = 0.058$), which marginally fails to meet the Bonferroni-corrected significance threshold of $\alpha=0.016$. This finding contrasts sharply with \citet{chattopadhyay2020tale}'s results, where biased actions were highly associated with reversals (p-value $<0.05$). In our LLM-assisted development context, only 27.08\% (257/949) of biased actions resulted in reversals, compared to their finding of 79.64\% (759/953). This raised some questions \modified{such as}: \emph{Does programming with LLMs reduce the need for reversing actions? }

\subsection{\modified{LLM Influence on Biases and Reversals}}

\begin{figure}
    \centering
    \includegraphics[width=\linewidth]{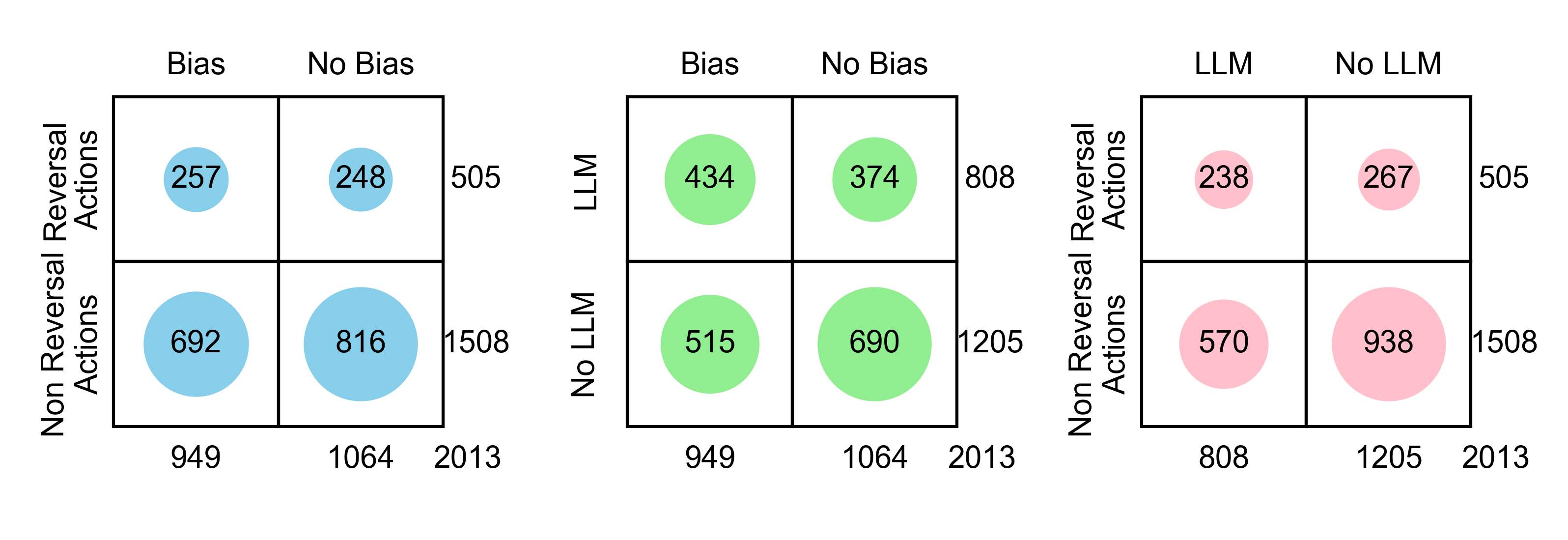}
    \caption{Distribution of Presence of (a) Bias and Reversal Actions, (b) Bias and LLM actions, and (c) LLM and Reversal actions. Circle size reflects action counts, and each cell shows actions matching that pair. Marginal totals appear along the bottom and right, with overall totals in the lower-right corner.}
    \label{fig:pre-bias-rev-llm}
    \Description{Distribution of Presence of (a) Bias and Reversal Actions, (b) Bias and LLM actions, and (c) LLM and Reversal actions. Circle size reflects action counts, and each cell shows actions matching that pair. Marginal totals appear along the bottom and right, with overall totals in the lower-right corner.}
\end{figure}

As participants used LLMs in our study, we wanted to better understand the role of LLMs in influencing cognitive biases. As described in the methodology~\Cref{action-coding}, we identified participant action with and without LLM. Figure~\ref{fig:pre-bias-rev-llm} (b) shows the proportion of actions associated with biases and LLMs. A chi-square test of independence revealed a highly significant association between LLM usage and the presence of cognitive biases ($\chi^2 (1, 2013) = 22.94, p = 1.67e-06$). Specifically, 53.71\% (434/808) of LLM-related actions were associated with at least one bias category. This suggests that using LLMs increases the likelihood of making biased actions during development. The higher prevalence of biases in LLM-actions raises the question: \emph{Are LLMs creating new cognitive blind spots? Do these tools influence over-reliance on generated suggestions while diminishing developers' critical evaluation skills?} To further investigate this phenomenon, we examined whether LLM usage directly influences reversal actions. Figure~\ref{fig:pre-bias-rev-llm} (c) presents the distribution of actions associated with LLMs and reversal actions. A chi-square test comparing LLM actions against reversal actions revealed another significant association ($\chi^2 (1, 2013)=13.32, p = 0.0003$). LLM-related actions had a reversal rate of 29.46\% (238/808).

Combined with our previous findings, \textit{a paradox emerges}: LLM actions are both more likely to contain cognitive biases (53.71\%) and more likely to be reversed (29.46\%). However, this creates a contradiction---then why did we not find a significant association between biases and reversals in our earlier analysis? This disconnect suggests that the prior $preCB$ bias categories may not fully explain the reversal patterns we observe in LLM-assisted development. The higher reversal rates in LLM actions, despite the weak bias-reversal correlation, indicate that there may be additional cognitive biases at play---ones that are unique to human-LLM interaction or are inadequately captured by the prior bias categorization. This points to the need for identifying new bias categories that better explain patterns in LLM-assisted software development.



Recent research on cognitive offloading to AI shows that it can reduce critical thinking and increase vulnerability to cognitive biases~\cite{Gerlich2025AITI}. Additionally, studies on biases in human–LLM interactions have identified types of bias that prior research has not addressed (such as Group Attribution Bias).  This may help explain some of the gaps we observed, as relying on LLMs could introduce new forms of bias. These findings support the need to expand our analysis to include a broader range of biases.

\section{Bias Categories \modified{in} LLM-Assisted Development}
\label{sec:rq1}
\begin{table*}[!th]
  \scriptsize
  \caption{Cognitive Bias Categories and included Biases \modified{(CBs marked with $^*$ were not observed in our study)} }
  \label{bias-categorizations}
  \def\arraystretch{1} 
  \resizebox{\linewidth}{!}{
  \begin{tabular}{p{0.3cm}p{1.8cm}p{6.2cm}p{7.5cm}}
    \toprule
     & \textbf{Category} & \textbf{Definition} & \textbf{Sample Biases} \\
    \midrule
    CB1 & \CBone & Select options or interpret information based on prior beliefs & Confirmation Bias~\cite{mahoney1977publication}, False Priors~\cite{pennycook2018prior}, Backfire Effect~\cite{sanna2002debiasing}, Semmelweis Reflex~\cite{mortell2013physician}\\ \hline
    CB2 & \CBtwo & Prefer suggestions from a more systematic or automated party without evidence of better performance & Automation Bias~\cite{goddard2011automation}, Authority Bias~\cite{milgram1963behavioral}, Halo Effect~\cite{nisbett1977halo}, Reactive Devaluation~\cite{ross1991barriers} \\ \hline
    CB3 & \CBthree & Prefer familiar approaches or tools even when their performance is suboptimal & Law of the Instrument~\cite{moore1983archaeology}, Well-travelled Road Effect~\cite{jackson1982empirical}, Mere Exposure Effect~\cite{zajonc2001mere} \\ \hline
    CB4 & \CBfour & Over-attribute group features to individuals or vice versa and allow that impression to influence decision & Group Attribution Error~\cite{allison1985group}, Stereotyping~\cite{correll2002police}, Ingroup Bias~\cite{taylor1981self}, Unconscious Bias~\cite{oberai2018unconscious}\\ \hline
    CB5 & \CBfive & Prefer to choose or defend self-created artifacts or self-maintained judgments even in the presence of better alternates & Endowment Effect~\cite{morewedge2015explanations}, Effort Justification~\cite{norton2012ikea}, Egocentric Bias~\cite{ross1979egocentric}, Illusion of Validity~\cite{tversky1974judgment}, Hindsight Bias~\cite{pohl2004cognitive}
    \\ \hline
    CB6 & \CBsix & Anchor efforts on initial assumptions even with added information or contradictory evidence or changed scope & Anchoring Bias~\cite{furnham2011literature}, Functional Fixedness~\cite{german2005functional}, Plan Continuation Bias~\cite{tuccio2011heuristics}, Scope Neglect~\cite{kahneman2000evaluation}, Sunk Cost Fallacy~\cite{staw1976knee} \\ \hline
    CB7$^*$ & \CBseven & Prefer default options and not seek alternatives even if the defaults are not optimal & Default Effect~\cite{herrmann2011effect}, Status-quo Bias~\cite{samuelson1988status}\\ \hline
    CB8 & \CBeight & Overestimate/Underestimate their capabilities, or the possibility of a good outcome or feel overly positive/negative about new technology & Optimism Bias~\cite{sharot2011optimism}, Planning Fallacy~\cite{buehler1994exploring}, Normalcy Bias~\cite{drabek2012human}
    , Pro-innovation Bias~\cite{rogers2014diffusion}, Pessimism Bias~\cite{sharot2007neural} \\ \hline
    CB9 & \CBnine & Prefer a solution/approach that provides instant gratification or a feeling of social belonging even when it will introduce errors in the long run & Hyperbolic Discounting~\cite{thaler1981some}, Present Bias~\cite{chakraborty2021present}, Outcome Bias~\cite{baron1988outcome}, Bandwagon Effect~\cite{nadeau1993new} \\ \hline
    CB10 & \CBten & Irrationally focus on one perspective of an event, ignoring other perspectives or possibilities & Survivorship Bias~\cite{ioannidis2005most}, Domain Neglect Bias~\cite{mike2022common}, Neglect of Probability~\cite{sunstein2002probability}, Salience Bias~\cite{schenk2011exploiting}, 
    \\ \hline
    CB11 & \CBeleven & Altered perceptions of options when compared, different from when assessed individually & Contrast Effect~\cite{plous1993psychology}, Framing Effect~\cite{tversky1981framing}, Less-is-better Effect~\cite{hsee1998less}, Distinction Bias~\cite{hsee2004distinction} \\ \hline
    CB12 & \CBtwelve & React based on the most recent or most readily available information in memory & Primacy Effect~\cite{glenberg1980two}, Recency Effect~\cite{baddeley1993recency}, Euphoric Recall~\cite{neighbors2019cognitive}, Memory Inhibition~\cite{macleod2007concept}, Google Effect~\cite{sparrow2011google}
    \\ \hline
    CB13$^*$  & \CBthirteen & Believe there is something meaningful behind data/ events when there is not & Apophenia~\cite{conrad1958onset}, Illusory Correlation~\cite{chapman1969illusory}, Agent Detection Bias~\cite{gray2010blaming}\\ \hline
    CB14 & \CBfourteen & View AI/tools as possessing human traits, emotions, or intentions & Anthropomorphic Bias~\cite{waytz2010sees}, Anthropocentric Thinking~\cite{coley2012common}, Truth Bias~\cite{millar1997effects}, 
    \\ \hline
    CB15 & \CBfifteen & Make risk-averse decisions that prioritize certainty over optimal outcomes & Ambiguity Effect~\cite{frisch1988ambiguity}, Zero-risk Bias~\cite{baron1988outcome}, Risk Compensation~\cite{hedlund2000risky} \\
    \bottomrule
  \end{tabular}}
\end{table*}







\modified{We categorized 90 relevant biases into 15 categories through a negotiated agreement with cognitive science experts.} \Cref{bias-categorizations} summarizes the 15 categories, sample biases, and their definitions (a categorization list for all biases available in the supplementary).
In the following paragraphs, we walk through each category.


\CBone \space(CB1) biases influence how easily developers believe LLM information matches their prior beliefs. For example, when wanting to solve path-finding with Dijkstra's algorithm, developers \modified{prefer using the same algorithm without considering alternatives, which confirms their beliefs}. \CBtwo \space(CB2) biases make developers believe LLM responses are superior to human judgment due to the system's mathematical construction. They may ask complex technical questions and accept answers without reflection, despite being aware of LLM's limitations. \CBthree \space(CB3) biases \modified{incline developers toward familiar tools and approaches, leading them to reiterate prior prompting strategies instead of investigating alternative techniques}.


\CBfour \space(CB4) biases refer to generalizing individual features to group features or vice versa. For example, \modified{after being disappointed} by LLM's inability to resolve a bug, a participant avoided LLM for all debugging tasks. \CBfive \space(CB5) biases occur when developers overvalue their judgment or self-made artifacts. One participant preferred writing and fixing a complex query rather than asking LLM for a new one. \CBsix \space(CB6) biases happen when developers anchor subsequent actions on initial assumptions despite changed validity. \CBseven \space(CB7) biases refer to preferring default choices or maintaining current states over alternatives. \CBeight \space(CB8) biases\footnote{This category can also contain cognitive tendencies: Valence Effect, Wishful Thinking, and Overoptimism. These biases are not in \cref{bias-categorizations} as they are synonyms.} occur when developers over/underestimate outcomes or are overly optimistic/pessimistic about themselves or tools. Developers may use vague prompts due to overconfidence in the interpretive abilities of LLMs.

\CBnine \space(CB9) biases make developers favor immediate gratification despite long-term costs. Developers may accept LLM code without reflection because programs run initially, despite introducing avoidable issues. \CBten \space(CB10) biases occur when developers irrationally focus on one perspective while ignoring others, \modified{such as} focusing only on the first part of LLM's long instruction lists. \CBeleven \space(CB11) biases alter solution perception during comparisons. Developers may judge LLM code as exceptionally good versus novice programmers but inadequate versus senior engineers, distorting objective assessment of AI capabilities. \CBtwelve \space(CB12) biases affect information memory and choice impact. Developers may prefer LLM-suggested packages they remember using recently.

\CBthirteen \space(CB13) biases occur when developers falsely believe meaningful links exist. A developer might observe LLM generating error-prone Java versus Python code and erroneously conclude intentional design for Python superiority. \CBfourteen \space(CB14) biases describe viewing LLM tools as possessing human traits, emotions, or intentions. Developers often comment ``\textit{Great}'', ``\textit{Wonderful}'', or ``\textit{Thank you}'' when responding to LLM assistants. \CBfifteen \space(CB15) biases lead to risk-averse decisions prioritizing certainty over optimal outcomes. When using LLM code, developers may reject uncertain solutions for familiar but inefficient approaches.

\section{Presence and Impact of Biases}

\subsection{\modified{Bias Frequency} in LLM programming (RQ1)}
\label{sec:rq2}

\begin{figure}
    \centering
    \includegraphics[width=\linewidth]{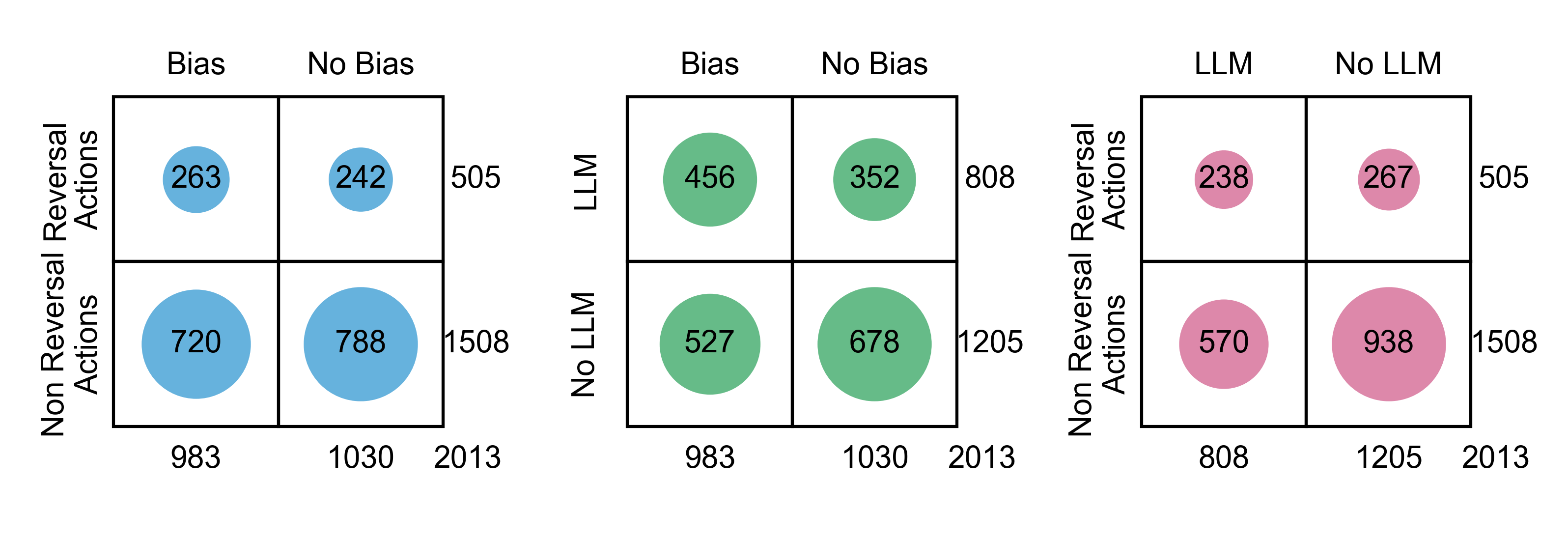}
    \caption{Distribution of Presence of (a) Bias and Reversal Actions, (b) Bias and LLM actions, and (c) LLM and Reversal actions. The size of the circles is proportional to the number of actions. Each cell presents the actions matching these dimensions.
Totals are shown along the bottom and right edges, with overall totals shown in the lower right-hand corners.}
    \Description{Distribution of Presence of (a) Bias and Reversal Actions, (b) Bias and LLM actions, and (c) LLM and Reversal actions. The size of the circles is proportional to the number of actions. Each cell presents the actions matching these dimensions. Totals are shown along the bottom and right edges, with overall totals shown in the lower right-hand corners.}
    \label{fig:newbiases-stats}
\end{figure}

We observed 2,013 actions across 14 participants: 808 (40.2\%) involved LLM interactions and 983 (48.8\%) were linked to cognitive biases. Among LLM actions, 456 out of 808 (56.4\%) involved at least one cognitive bias, with 393 out of 983 (40\%) biased actions involving multiple biases (\Cref{fig:newbiases-stats}).
We analyzed reversal actions following \citet{chattopadhyay2020tale}, finding 505 total reversed actions (25\%). LLM interactions showed higher reversal rates: 238 out of 808 actions (29.5\%), and among biased LLM actions, the reversal rate was 29.8\%.

Chi-square tests with Bonferroni correction revealed significant associations. LLM use strongly correlates with bias presence ($\chi2(1,N=2013) = 30.72, p=2.97 \times10^{-8}$), indicating LLM interactions are more likely biased. LLM interactions are also more likely to be reversed (($\chi^2(1, N = 2013) = 13.32, p = 2.62 \times10^{-4} $)). However, within LLM interactions, only \CBsix (($\chi^2(1, N = 808) = 7.15, p = 7.49 \times10^{-3} $)) and \CBthree ($\chi^2(1, N = 505) = 11.87, p = 5.67 \times10^{-4} $) showed significant bias-reversal associations. Our findings reveal that LLM-assisted programming increases both bias likelihood and reversal rates, with more than 50\% LLM interactions likely to be associated with biases.

\subsection{Bias Distribution across LLM-actions (RQ2)}

\begin{figure}[h]
  \centering
  \includegraphics[width=\linewidth]{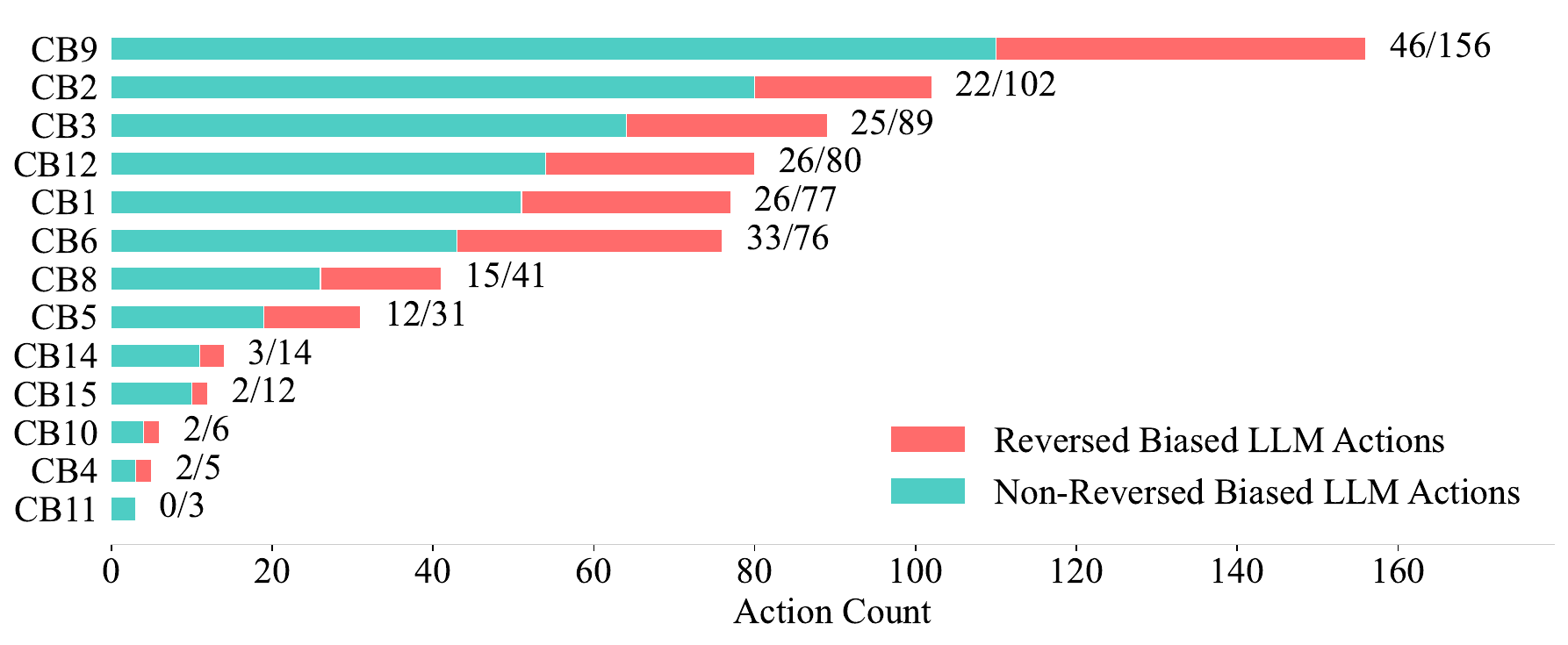}
  \caption{
  \modified{Frequency of biased LLM actions and reversals by category.}
  }
  \Description{Frequency of biased LLM actions and reversals by category.}
\label{fig:bias_dist}
\end{figure}

\modified{Given the significant association between LLM actions and bias presence, we examined how these biases manifested in LLM-related actions. \Cref{fig:bias_dist} presents bias distributions in LLM interactions and their corresponding reversal rates.}

\modified{\textbf{\CBnine (CB9)} was the most frequent bias in 156 LLM-related actions, with 46 reversals (29.5\%) across 13 participants. It stems from prioritizing speed over solution quality. P1 copied ChatGPT-generated CSS despite of being uncertained \textit{``[was] not sure [it] will work''}, which overwrote existing CSS and broke the UI, leading him to use ChatGPT again to restore it and producing a full sequence of reversed actions.}

\textbf{\CBtwo (CB2)} was second most frequent with 102 LLM-related actions, 22 reversed (21.5\%), in 10 participants. Participants adopted LLM suggestions without questioning, especially for complex tasks. P13, building a Flask dashboard, said he \textit{``did not want to deal with HTML''} and asked ChatGPT for a complete Python solution, copying without review. When it failed, he switched to Phind, repeating the same blind-pasting approach until finding an error-free version after seven minutes.



\textbf{\CBthree (CB3)} appeared in 89 actions, with 25 reversed (28\%), affecting seven participants [P2–4, 7, 10–13]. Developers favored LLM suggestions matching their existing tool knowledge. P4 asked ChatGPT to convert environment variables into Python variables—despite earlier requesting Google Cloud SDK guidance—due to greater Python familiarity. Though not reversed, this added nearly two minutes to a simple task, often leading to over-reliance on LLMs and limited critical assessment.

\textbf{\CBtwelve (CB12)} biases were associated with 80 developer-LLM actions, 26 reversed (32.5\%), affecting eight participants [P1,4, 6, 7, 9, 10, 11, 14]. Participants tried LLM solutions similar to recent experiences without considering appropriateness. P6 refactored a function using GitHub Copilot, initially accepting its suggestion: \textit{``Yeah, [it's] a correct suggestion.''} When encountering an error, she realized the approach was incorrect: \textit{``I just remembered. This suggestion kinda threw me off.''} This bias led to action reversal and implementing the function herself.

\modified{\textbf{\CBone (CB1)} biases occurred in 77 LLM-related actions, 26 reversed (33.8\%), affecting eight participants [P4, 5, 8-13]. Developers often adopted LLM suggestions based on preconceptions. P10 chose ChatGPT's \texttt{Chart.js} suggestion because \textit{``I had a library in mind using Chart.js, which I think is pretty popular for generating charts in the UI.''} False priors also led participants to accept or reject suggestions inappropriately. An initial faulty ChatGPT debugging response biased P8 against later suggestions; when ChatGPT finally returned the correct answer, he dismissed it as \textit{``It seems like it's just providing the same thing again.''}}

\textbf{\CBsix (CB6)} biases emerged in 76 LLM interactions, with 33 reversed (43.4\%), affecting seven participants [P1, 2, 4, 8, 9, 12, 14]. Developers anchored on initial solutions despite contradictory evidence. \modified{P12 copy-pasted ChatGPT’s file-reading code in Google Colab and hit an error. She repeatedly duplicated the same cell, attributing the issue to the environment, until she finally realized the cause was an incorrect file path.}

\textbf{\CBeight (CB8)} occurred in 41 LLM actions, with 15 reversed (36.6\%), affecting six participants [P1, 3, 4, 9-11]. Participants provided incomplete context prompts, expecting correct answers based on prior successes. P11 asked ChatGPT to add an \textit{edit} button to his HTML/CSS. After seeing the result with incorrect spacing, he assumed ChatGPT would know his spacing preferences from previous interactions, despite not specifying this in his prompt.


\textbf{\CBfive (CB5)} biases influenced 31 actions, 12 reversed (38.7\%), observed in five participants [P3, 6-8, 14]. They prioritized their own thoughts/code over LLM suggestions. P8 asked ChatGPT for troubleshooting help but dismissed the response, relying on incorrect assumptions instead. \textbf{\CBfourteen (CB14)} appeared in 14 actions, three reversed (21.4\%), involving seven participants [P1,2,6,7,10,11,14]. Participants attributed human-like abilities to LLMs, such as memory [P7], social awareness (\textit{``Thank you''} [P10], \textit{``Please''} [P1]), or inferring intention [P11]. P4 said \textit{``Copilot just knows what [he] wants to type,''} though the suggestion was incorrect.

\modified{Twelve actions involved \textbf{\CBfifteen (CB15)}, with two reversed (16.6\%) across four participants [P3,5,8,14]. P14 asked ChatGPT to generate a long CREATE TABLE query only to avoid errors. Others sometimes ignored LLM suggestions for the same reason.}

The remaining three categories emerged only occasionally, with fewer than 10 LLM-related actions associated with them. P11 encountered \CBten (CB10), where he repeatedly focused on only the functionality suggestions while ignoring the aesthetics.
Participants influenced by \CBfour (CB4) generalized one specific LLM failure instance in one problem to other interactions. Only P2 encountered \CBeleven (CB11); despite being satisfied initially, he lost confidence when two LLM suggestions were incorrect.

\section{\modified{Managing Consequences of Biases (RQ3)} }
\label{sec:mitigations}


To gauge whether developers believe biases manifest in programmer-LLM interactions, we analyzed survey responses from 22 participants. We asked \textit{``How often do you think software developers experience [bias category] scenarios?''} and \textit{``How do such [bias category] scenarios impact the software development process''}. Scenarios captured instances from our observed data of participants experiencing biases when programming with LLMs.



\begin{figure*}
    \centering
    \begin{subfigure}[t]{0.48\linewidth}
        \centering
        \includegraphics[width=\linewidth]{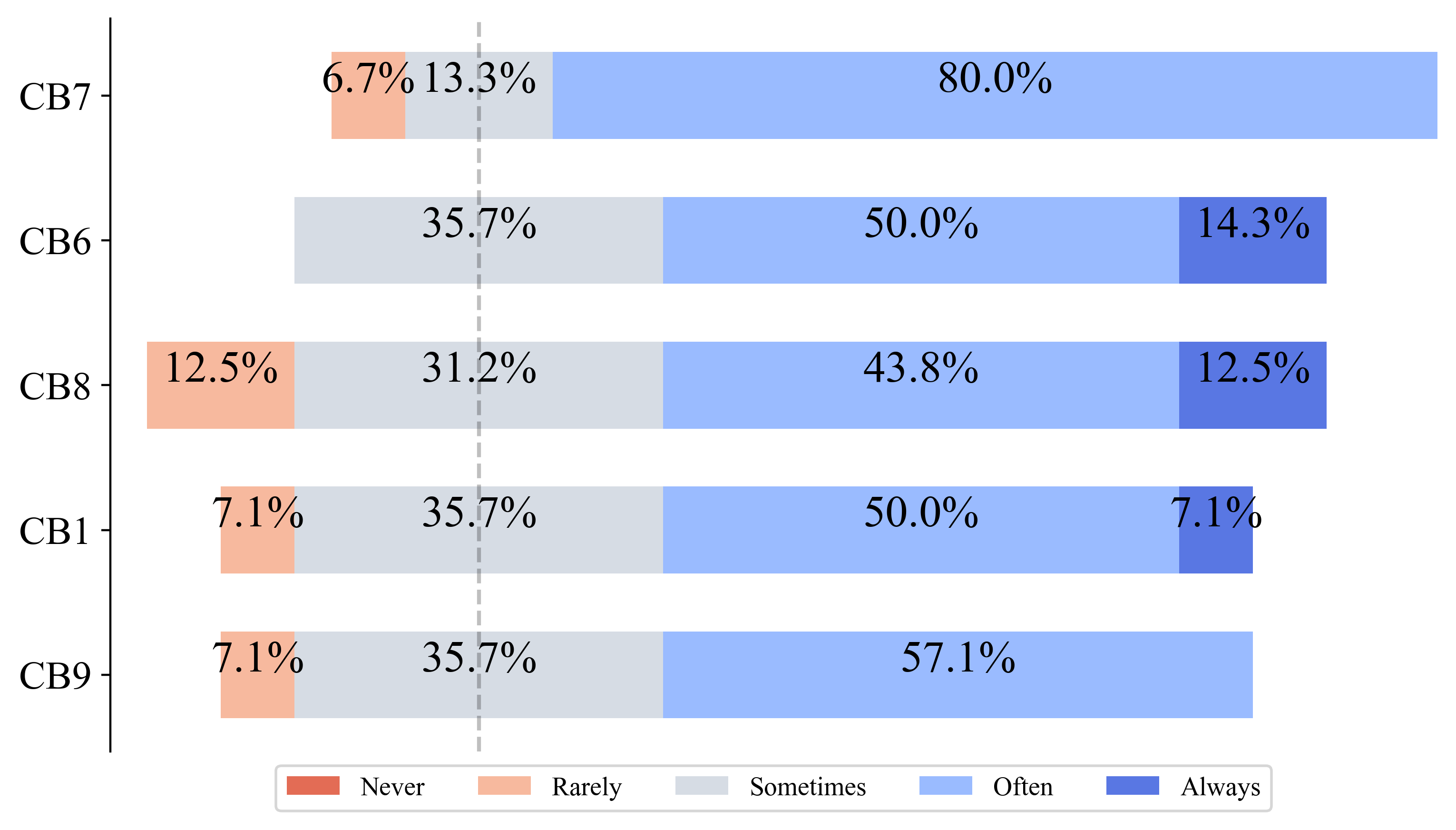}
    \end{subfigure}
    \hfill
        \begin{subfigure}[t]{0.48\linewidth}
        \centering
        \includegraphics[width=\linewidth]{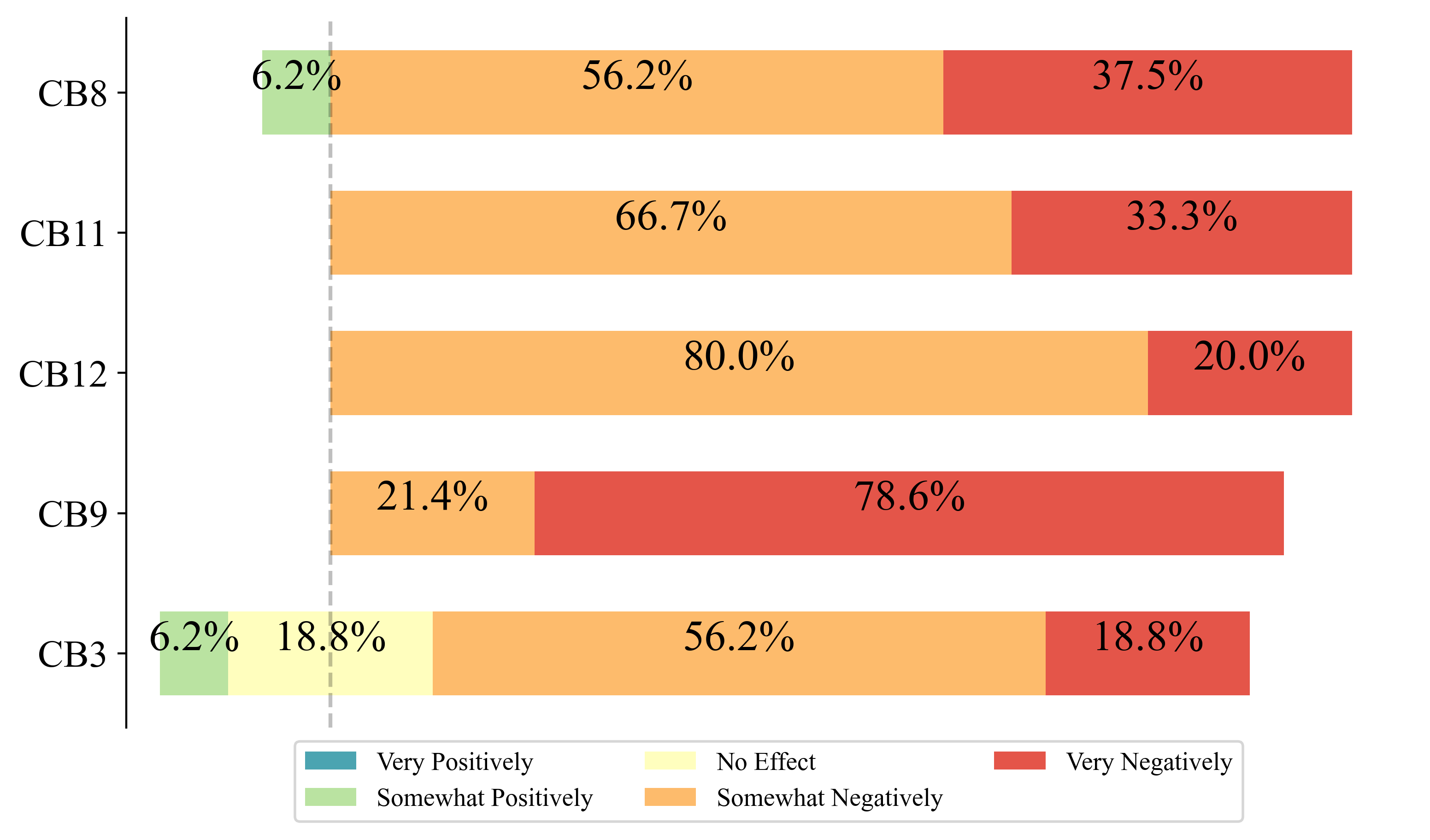}
    \end{subfigure}
  \caption{ (a) The five highest perceived frequencies of bias. Bars are ordered by the combined proportion of ``Often'' and ``Always'', with dark + light blue segments indicating high-frequency, red indicating low-frequency responses (``Never'', ``Rarely''); center grey segment represents neutral ``Sometimes''. (b) The five highest perceived impactful biases. Bars are sorted based on the sum of the ``Somewhat negative'' and ``Very negative'' frequencies with red bars, while green bars signify positive impacts, with bars in the center reflecting ``No effect'' responses.}
  \label{fig:survey-q1q2}
  \Description{ (a) The five highest perceived frequencies of bias. Bars are ordered by the combined proportion of ``Often'' and ``Always'', with dark + light blue segments indicating high-frequency, red indicating low-frequency responses (``Never'', ``Rarely''); center grey segment represents neutral ``Sometimes''. (b) The five highest perceived impactful biases. Bars are sorted based on the sum of the ``Somewhat negative'' and ``Very negative'' frequencies with red bars, while green bars signify positive impacts, with bars in the center reflecting ``No effect'' responses.}
\end{figure*}

\Cref{fig:survey-q1q2}-a shows perceived top five highest frequency distributions ranging from ``Never'' to ``Always''. Survey participants are aware that biases frequently occur in programmer-LLM interactions. No cognitive bias was reported as ``never'' occurring. Eight bias categories (CB7, CB6, CB8, CB1, CB9, CB13, CB4, CB5) were perceived to occur \modified{``often'' or ``always''} more than 50\% of the time.

\Cref{fig:survey-q1q2}-b shows top five highest perceived impact ranging from Very \modified{Positively to Very Negatively}. The percentage of mentions is shown on the respective sections of the bars. Participants are aware of the negative impacts, as indicated by the high proportions of red bars. \CBfourteen (CB14) was the only category where participants were divided, with 57\% reporting no impact, which is reasonable since empathy in LLM prompts likely won't change outputs. For all other categories, at least 50\% of participants reported negative impacts. The complete plot is in the supplementary.

While cognitive biases occur frequently and negatively impact programming, they are often challenging to detect in a timely manner. 
\modified{Differences between observed bias impacts and survey rankings highlight a gap between actual effects and developer perceptions, making bias avoidance difficult without systematic support. We discuss participant-suggested practices and tools for mitigation.}


\subsection{Helpful practices}

Two authors identified 18 categories of helpful practices, which were later consolidated into four broader categories based on when they occur in the programming process with LLMs. \Cref{tab:helpful_practices_1} shows these categories, subcategories, and the related bias survey section.

\begin{table}[ht]
\footnotesize
\caption{\modified{Helpful Practices Suggested by Participants with Descriptions and Targeted CBs.}}
  \def\arraystretch{1} 
  \resizebox{\linewidth}{!}{
\begin{tabular}{|p{1.2cm}|p{6.1cm}|p{0.7cm}|}
\hline
\textbf{Categories}                                 & \textbf{Sub-categories}                                                                                                                                     & \textbf{Target CB}           \\
\hline
\multirow{6}{=}{Before Prompting}              & \textbf{Requirement Analysis:} planning, clarify requirements                                                                                               &  3, 4, 14     \\
                                            & \textbf{Knowledge Preparation:} Understanding previous architecture choices,  current frameworks/technology/library, new technology/library, sample implementation       &  3, 4, 11, 12   \\
                                            & \textbf{Break Down Tasks:} Problem formation, break down task, rapid iteration and prototyping                                   &  2, 4,5,9, 11, 13   \\
                                            & \textbf{System Design:} what frameworks/library/technology to use, weigh between familiar and trending, modular design                         &  7, 15   \\
                                            & \textbf{Test-Driven Development:} come up test first according to requirement, preventive testing and error handling                               &  9, 10-12, 15   \\
                                            & \textbf{AI As A Tool:} AI has it's own limitation, balance LLM use with your own coding expertise when collaborating with AI, code independently before asking LLM  &  2, 13, 14   \\
\hline
\multirow{2}{=}{Prompting}                  & \textbf{Evaluation Help:} let LLM go over the logic to find issues early, let LLM explain its suggestions                                            &  5, 8, 13    \\
                                            
                                            & \textbf{Prompt Writing:} precise, well formulated (formatted), more context, add examples, structured with steps                                                       &  11-15   \\
\hline
\multirow{7}{=}{Evaluating LLM Responses}   & \textbf{Understand Before Adopting:} use flow chart / read carefully to understand the code logic line by line                                              &  6, 11    \\
                                            & \textbf{Code Quality Matters:} quality over speed & 5, 9   \\
                                            & \textbf{Incremental Development:} develop by small pieces incrementally and run test every step                                     &  2-11, 13-15   \\
                                            & \textbf{Open-minded:} unfamiliar, unexpected solutions; use as inspiration or starting point, willing to revisit previously rejected LLM suggestions        &  1, 2, 4, 5, 10, 11, 13   \\
                                            & \textbf{Seek Alternatives:} test all hypotheses, test all alternatives before adopting (across LLMs, across responses), side-by-side examination            &  1, 2, 6, 7, 11    \\
                                            & \textbf{External Resources:} reference guidelines,  cross-checking for accuracy / truthfulness                                          &  2, 4, 6, 11, 13   \\
                                            & \textbf{Peer Support:} peer code review, knowledge sharing, pair programming                                                                                &  1-15  \\
\hline
\multirow{5}{=}{After Applying LLM Suggestions} & \textbf{Long-term Impact:} test case generation, review test coverage report, regularly run regression tests                                             &  12, 15 \\
                                            & \textbf{Code Maintenance:} document changes, managing module dependencies                                                                                   &  7, 9, 15   \\
                                            & \textbf{Able to Revert:} version control and feature flags                                                                                                  &  7, 15   \\
\hline
\end{tabular}
}
\label{tab:helpful_practices_1}
\end{table}

\textbf{Before Prompting.} In addition to understanding requirements and making design decisions, participants emphasized Test-Driven Development (TDD) and maintaining an ``AI as a tool'' mindset. TDD, which alternates between writing failing tests and implementing code~\cite{nagappan2008realizing}, helps manage risks and verify changes, reducing biases [SP22] such as those found in \CBnine, \CBten, and \CBfifteen. Viewing LLMs as limited tools, avoiding blind affirmation [SP3] and keeping interactions functional [SP6], supports objectivity and mitigates \CBtwo, \CBthirteen, and \CBfourteen. Developers also noted the importance of trying to solve issues independently before consulting LLMs [SP11] and balancing tool use with their own expertise [SP10].

\textbf{Prompting.} In addition to debugging and implementation, participants noted asking LLMs for early feedback to avoid wasted effort [SP3,4,10], helping mitigate \CBfive\ and \CBeight. SP5 advised requiring LLMs to explain their suggestions to catch misunderstandings (\CBthirteen). Participants highlighted the value of clear, structured prompts with detailed context. SP11 recommended ``formulating questions with clear, logical steps.''

\textbf{Evaluating LLM Responses.}
Evaluation is central to effective developer–LLM interaction, and relies on understanding suggestions. SP21 emphasized that ``process flow charts and analyzing code line by line'' help build that understanding. 14 participants highlighted that incremental development, such as TDD, could be beneficial. SP19 advised to ``refactor as you go'' for cleaner evaluations. Despite the speed of LLM responses, participants cautioned against sacrificing quality, particularly under \CBnine, and stressed ``prioritizing code quality over speed'' [SP14]. Participants also recommended debugging LLM outputs [SP6, 10, 19], comparing across models [SP3], and validating hypotheses [SP9].


\textbf{After Applying LLM Suggestions.}
Adopting the LLM suggested code requires measures to maintain system stability. Regression tests run after integrating LLM suggestions can ensure that the ``functionality remains intact [SP6]'' (\CBfifteen). Documentation must be kept up to date to ensure that unnecessary changes are avoided by LLMs [SP12,15,22]. Also, reversible changes can encourage developers to try out alternative (\CBseven) and relatively risky LLM suggestions (\CBfifteen) by utilizing version control and feature toggle flags.

\subsection{Helpful tools}
\label{sec:mitigations-tools}

13/22 \pts provided useful tool suggestions for some CBs.

\textbf{\CBtwo (CB2)} negative effects can be reduced by using code refactoring tools such as IntelliJ and VScode [SP9,11,12], which fix any errors that might be overlooked by a false impression of LLM suggestions. \textbf{\CBthree (CB3)} bias can be mitigated with tools that expose developers to alternate solutions, increasing their range of possible options. SP4 and SP19 mentioned using ``Framework Comparison Tools such as Libscore and Dependabot''. \textbf{\CBeight (CB8)} negative outcomes, specifically potential bugs and errors in LLM-generated code, can be mitigated using error-tracking tools such as Sentry or LogRocket [P19] and Continuous Integration/Continuous Deployment (CI/CD) setups (e.g., Jenkins, CircleCI), which ``can automatically run tests on new code before it gets merged'' [P9]. \textbf{\CBeleven (CB11)} bias might cause developers to perceive the quality of LLM-generated code as inferior to that of online examples or demos. To address this, \pts [SP6,10,18,22] suggested using code comparison tools that help developers make more informed decisions about whether to use LLM-generated code, like Beyond Compare, DiffMerge, or WinMerge. P8 suggested that \textbf{\CBthirteen (CB13)} can be mitigated by using tools ``that provide explanations behind LLM suggestions to help developers understand the logic and context.''

\subsection{LLM improvement suggestions}

One author thematically organized LLM improvement suggestions into five groups, which were vetted by an NLP researcher specializing in code-focused LLMs.

\textbf{Explicit Reasoning Augmented Suggestions:} Participants suggested that LLMs providing detailed reasoning could mitigate seven of 15 biases. Step-by-step explanations help compare preconceptions with logic (\CBone, CB1) [SP8, 16, 19], shift focus from \textit{who} to \textit{why} (\CBtwo, CB2) [SP1, 5, 8, 11, 18], make unfamiliar options accessible (\CBthree, CB3) [SP6, 21, 22], and improve transparency (\CBfour, CB4) [SP10, 13-15]. Trade-off explanations guide prioritization, mitigating \CBfive (CB5) and \CBthirteen (CB13) [SP3, 6, 8, 14-16, 18, 2].



\textbf{Solution Diversity in Suggestions:} Participants suggested that LLMs presenting multiple alternative solutions could mitigate two CBs. This approach helps with \CBone (CB1) by compelling developers to compare different approaches rather than defaulting to preconceptions [SP4,12-14]. It also addresses \CBsix (CB6) by breaking rigid initial assumptions when developers get stuck on a single approach [SP4,8,9,19].

\textbf{Context-Aware Suggestions.}
Participants recommended improving LLMs’ awareness of project context to mitigate four cognitive biases. For \CBone, aligning suggestions with developers' style [SP6] and prior interactions [SP9] aids comprehension. For \CBthree, recommending frameworks suited to developer expertise [SP10] and project needs [SP9] prevents bias toward familiar tools. For \CBfour, understanding requirements [SP5] and code context [SP19] avoids overgeneralization. For \CBseven, suggesting libraries and handling dependencies [SP11, SP22] reduces effort in choosing suitable solutions.

\textbf{Automated Proactive Verification of Suggestions.}
Participants proposed that LLMs reduce bias through improved error handling and verification. For \CBsix, they suggested error logging [SP9], contextualized explanations [SP11], and error categorization [SP12] to encourage exploring alternatives. To address \CBeight, they recommended integrated testing [SP3, SP5, SP19], warnings [SP22], real-time error detection [SP9], and flagging undefined functions [SP11]. For \CBnine, LLMs should warn about likely errors [SP10, SP13] to discourage quick, low-effort choices. Encouraging self-reviews.

\textbf{Interactive Dialogues Pre-suggestions:} Participants wished for more interaction. Asking for clarification when users repeatedly return with the same issue [SP21] can break fixation (\CBsix (CB6)). For \CBtwelve (CB12), LLMs should request further input [SP13] for comprehensive context [SP3,15] and ask clarifying questions for each assumption [SP20]. Participants also recommended more conversational interactions [SP13] and clear communication about LLM limitations [SP3,15] to ensure developers don't perceive AI as more capable or human-like than it is \CBfourteen (CB14).


\section{Discussion}

We observed that cognitive biases were related to \pts' attitude towards LLM and on how they use LLMs. 

\subsection{Biases and Attitude towards LLMs}
\begin{figure}
    \centering
    \includegraphics[width=\linewidth]{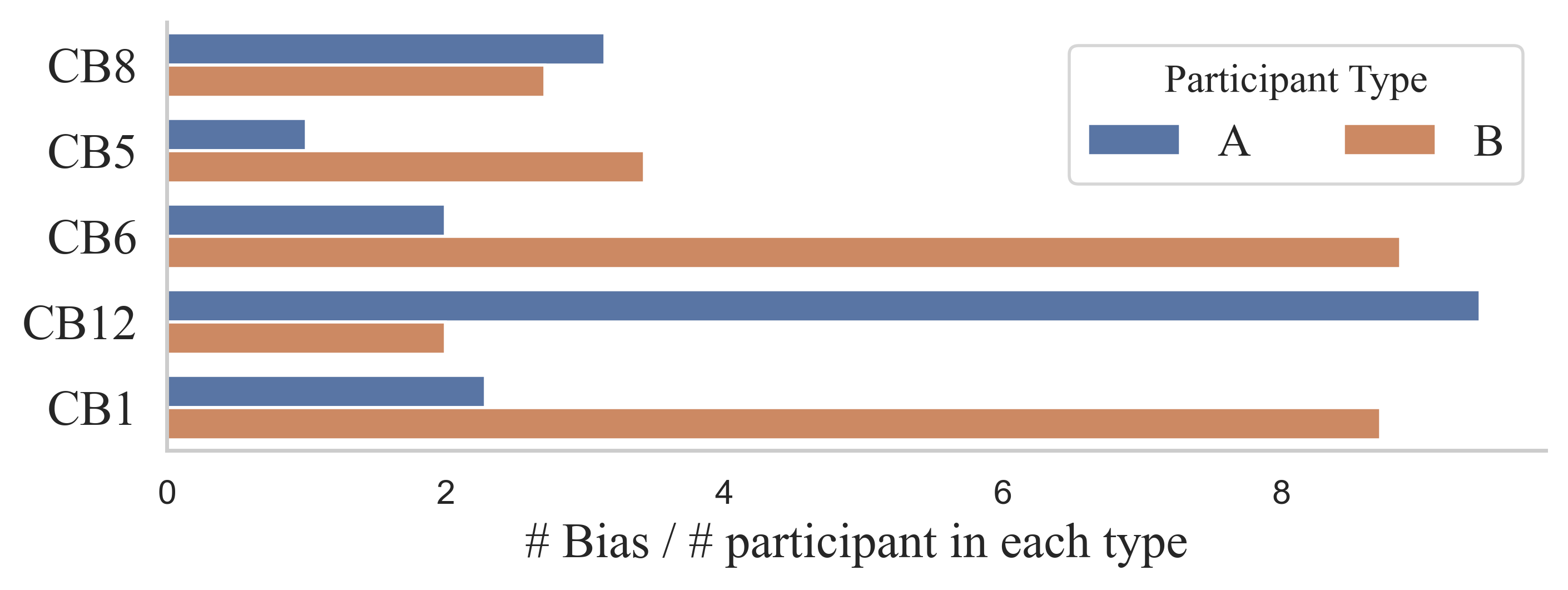}
     \caption{Bias categories differing between the two groups. Counts are normalized by the number of participants in each group.}
     \Description{Bias categories differing between the two groups. Counts are normalized by the number of participants in each group.}
    \label{fig:biases_per_part_type}
\end{figure}

\modified{Thematic analysis conducted by two researchers revealed two task delegation attitudes toward LLMs. Each quotation was coded as Type A or B; participants were labeled by their majority strategy. \Cref{fig:biases_per_part_type}-a maps delegation strategies to Type A (P2,4,5,7,9,11,12) or Type B (P1,3,6,8,10,13,14).} \typeA participants relied on LLMs for high mentally demanding tasks like ``going through documentation and understanding every parameter'' [P2], ``debugging errors by directly prompting the LLM'' [P12], or ``brainstorming new ideas'' [P5], aligning with exploration-phase programming \cite{barke2023grounded}. 
Conversely, \typeB participants delegated repetitive or low mentally demanding tasks like ``figuring out correct package import format'' [P1], ``generating 50 similar query entries'' [P14], or ``coding a loop in this language'' [P10], reflecting acceleration-phase automation \cite{barke2023grounded}. Additionally, seven participants [P2,3,5-7,9,11] mentioned using ``LLM to get template solutions as starting points''. We calculated the frequency of biases associated with participants observed from each type.
\Cref{fig:biases_per_part_type} shows the bias categories that differ significantly.



\emph{\typeA.} Developers who delegate \edited{high mentally demanding tasks to the LLM often relinquish control due to optimism about LLMs' capabilities (CB8) and expect it to provide correct answers}. For example, \new{P11 asked the LLM to set up a React website, receiving a long step-by-step plan. They completed the first step but missed the second, causing errors. Delegating the complex task led to a lengthy LLM response, making it hard to recall all details (CB12).}



\emph{\typeB.} \modified{Developers who tackle high mentally demanding tasks themselves and delegate easier ones to LLMs, preferring control (CB5) but often persisting rigidly (CB6). For example, P13, biased against using HTML (CB1), struggled to make a non-HTML approach work (CB5) but failed.}

\subsection{\modified{Bias Across Task types}}

In this study, participants engaged in their routine development projects and completed different tasks. We categorized their task based on the developmental complexity of the work they engaged in during each session. Most participants were involved in coding, debugging, and system implementation, and we classified them into three groups based on the intensity and scope of these activities: heavy, medium, and low development effort. We observed that participants working on heavy development tasks (P6, P9) engaged in more biased interactions with the LLM on average (mean = 112 instances). In contrast, participants performing medium-level tasks (e.g., translating data to English, retrieving datasets) or low-level tasks (e.g., reading or manipulating tabular data) exhibited comparable frequencies of bias, with no statistically significant difference between these two groups.

Developers facing more complex tasks appear to rely more heavily on LLM assistance. Higher cognitive load and time pressure likely reduce scrutiny of model outputs, leading to susceptibility to biased decisions. However, this observation is preliminary. Larger and more diverse samples are needed to robustly assess how task complexity shapes bias in LLM interactions ~\cite{Wang2019Designing}.

\section{\modified{Implications}}

Beyond implications for developers and tool builders (Section~\ref{sec:mitigations}), our work highlights directions for researchers and LLM developers.

\paragraph{Future Research in Mitigating Biases.} While our findings surfaced tools and practices that developers perceived as useful, further research needs to investigate interventions. Table~\ref{tab:literature_mitigations} presents various bias mitigation techniques drawn from multiple domains outside computer science that can be applied at different stages of interaction with LLMs. Prior work focuses on the ``Before Prompting'' stage, emphasizing general debiasing strategies that can be adapted to our context. Common approaches include mindset enhancement techniques such as pre-mortem analysis~\cite{Kinsey2020BurningBM} and devil’s advocacy~\cite{liden2019devil}.

\begin{table}[h!]
\scriptsize
\centering
\caption{Bias Mitigation Strategies from Other Domains}
\vspace{-0.5em}
\renewcommand{\arraystretch}{1.1}
\setlength{\tabcolsep}{3pt}
\begin{tabular}{|p{1.7cm}|p{6.4cm}|} 
\hline
\textbf{Category} & \textbf{Method Description and Source} \\ 
\hline
\textbf{Before Prompting} &
Pre-mortem checklists: anticipate failures and surface assumptions early. \cite{Kinsey2020BurningBM} (HBE) \\
\cline{2-2}&
Thematic-based categorization, stakeholder engagement, bias checklists, and continuous monitoring \cite{Hasanzadeh2025BiasRA} (HC) \\
\cline{2-2}
& Devil’s advocacy: assign roles to question early ideas. \cite{liden2019devil} (Psyc)\\
\cline{2-2}
& Outside-view planning: use reference cases to counter optimism. \cite{shmueli2016can} (Mgmt)\\
\cline{2-2}
& Task segmentation: divide work to reduce overconfidence. \cite{forsyth2008allocating} (Psyc,Mgmt) \\
\hline
\textbf{Prompting}
& Self-explanation: articulate reasoning to spot errors.\cite{10.1145/3710946} (HCI) \\
\cline{2-2}
& Extends choice architecture by adding a two-way feedback loop: LLM flags flaws, humans refine prompts \cite{rastogi2022deciding} (DecisionHAI) \\
\cline{2-2}
\hline
\textbf{Evaluating LLM Responses}
& Continuously monitoring AI systems to detect and address emerging biases over time \cite{fasolo2025mitigating} (Mgmt) \\
\cline{2-2}
& Cross-check with sources and verify factual claims. \cite{ke2024mitigating} (CL SC) \\
\cline{2-2}
\hline
\end{tabular}
\label{tab:literature_mitigations}
\parbox{0.99\linewidth}{%
  \tiny
\centering
 \new{Psyc: Psychology, Mgmt: Management, CL SC: Clinical Science, HBE: Human Behavior in Engineering, HC: Healthcare, HCI: Human Computer Interaction, DecisionHAI: Decision Making with Human AI} 
}

\end{table}


Few existing techniques directly target the prompting stage of interaction. Prior work has investigated the impact of changing deliberation time~\cite{rastogi2022deciding}, clarifying communication~\cite{eibl2025exploring}, and user education~\cite{sabouri2025trust} (aligning with our `Evaluation Help' and `Understand Before Adopting' strategies). Our findings in Table~\ref{tab:helpful_practices_1} also report proactive preparation phases preceding LLM use and strategies for `After Applying LLM Suggestions' that require further exploration.

\paragraph{\modified{Design Strategies for LLM Developers.}}
The speed and fluency of LLM-generated code can create an illusion of correctness, exposing developers to the risk of engaging with biases. While developers can implement practices themselves and rely on tools, some biases require interventions that encourage more careful evaluation of generated outputs. Prior work \citet{rastogi2022deciding} has shown that slowing down interactions can effectively mitigate certain cognitive biases. LLM designers can employ interaction methods or architectural changes that slow down the decision-making process.

LLM developers can introduce helpful friction into the workflow~\cite{inan2025better}. Adding productive friction, such as requiring explicit confirmation before integrating generated code, can promote more critical evaluation. For specific biases, such as anchoring bias, designers can interrupt the sequential verification process by periodically introducing alternative code analysis interventions, similar to \citet{echterhoff2022ai}. This helps break the chain of previously correct oversights that can reinforce the bias.
LLM builders could incorporate structured review prompts or checklists to guide systematic evaluation. Other design opportunities include adding time delays before code acceptance~\cite{liu2024ai}, or displaying confidence scores. For architectural changes, designers could implement ``slow thinking''~\cite{su2024dualformer}, which emphasizes System 2 reasoning in the Chain of Thought process to encourage more reflective user interaction~\cite{pareek2024trustdev}.
The output of such a model could prompt developers to think more critically about the code and understand the underlying rationale behind code generation, which has been shown to support better decision-making~\cite{cheng2019explaining}. Finally, while researchers have experimented with improving model architectures through fairness metrics and bias auditing~\cite{Gray2023Measurement,Shah2024A}, the impact of these changes on mitigating biases needs further investigation.




\section{Limitations}

\modified{Our study offers qualitative and quantitative insights into cognitive biases in developer-LLM interactions. With 14 observational and 22 survey participants, the sample size aligns with recent work~\cite{arteaga2024support}, though generalizability is limited.} To mitigate the limitations of participants having to work on unfamiliar projects during the session, \new{we instructed participants to work on a project of their own choice in a familiar, comfortable setting.} We also emphasized at the start of the study that we were there to ``learn'' from them. Additionally, participants' side projects may not fully capture the complexities of real-world products. Survey limitations include potential response bias, where participants may overestimate their awareness of biases or underreport negative impacts to appear more competent. \modified{The scenario-based approach, based on observed data, may not capture all bias complexity. We addressed this by analyzing cumulative perceptions of frequency and impact, rather than individual rankings.} \modified{Action reversals capture real-world bias indicators without disrupting behavior, but provide only a lower-bound estimate. Longitudinal studies are needed to assess longer-term effects.}





\section{Conclusion}


In this paper, we conducted a mixed-methods study and compared how developers encountered cognitive biases when working with LLMs versus the traditional non-LLM workflow. We found that working with LLMs exposes developers to new biases, and we developed a taxonomy of 90 biases that impact developer-LLM collaboration. We also observed that cognitive biases are related to \pts's attitude toward LLMs. Through a complementarity survey, we triangulated findings from the observation sessions and collected useful practices and tools that can mitigate the impact of biases in developer-LLM collaboration.

\begin{acks}
We are grateful to all the participants in our study. We also thank our collaborators, colleagues, and lab members who provided valuable feedback and support throughout the research.
\end{acks}
\newpage

\bibliographystyle{ACM-Reference-Format}
\bibliography{sample-base}

@String{Computing = "Computing" }

@String{Springer = "Springer-Verlag" }

@misc{github_2023_survey,
  author = {GitHub, Shani},
  month = {06},
  title = {Survey reveals AI’s impact on the developer experience},
  url = {https://github.blog/news-insights/research/survey-reveals-ais-impact-on-the-developer-experience/},
  urldate = {2025-07-19},
  year = {2023},
  organization = {The GitHub Blog}
}

@article{sanna2002debiasing,
  title={When debiasing backfires: accessible content and accessibility experiences in debiasing hindsight.},
  author={Sanna, Lawrence J and Schwarz, Norbert and Stocker, Shevaun L},
  journal={Journal of Experimental Psychology: Learning, Memory, and Cognition},
  volume={28},
  number={3},
  pages={497},
  year={2002},
  publisher={American Psychological Association}
}

@article{garrison2006revisiting,
  title={Revisiting methodological issues in transcript analysis: Negotiated coding and reliability},
  author={Garrison, D Randy and Cleveland-Innes, Martha and Koole, Marguerite and Kappelman, James},
  journal={The internet and higher education},
  volume={9},
  number={1},
  pages={1--8},
  year={2006},
  publisher={Elsevier}
}

@article{10.1145/3710946,
author = {de Jong, Sander and Paananen, Ville and Tag, Benjamin and van Berkel, Niels},
title = {Cognitive Forcing for Better Decision-Making: Reducing Overreliance on AI Systems Through Partial Explanations},
year = {2025},
issue_date = {May 2025},
publisher = {Association for Computing Machinery},
address = {New York, NY, USA},
volume = {9},
number = {2},
url = {https://doi.org/10.1145/3710946},
doi = {10.1145/3710946},
journal = {Proc. ACM Hum.-Comput. Interact.},
month = may,
articleno = {CSCW048},
numpages = {30}
}

@article{ke2024mitigating,
  title={Mitigating cognitive biases in clinical decision-making through multi-agent conversations using large language models: simulation study},
  author={Ke, Yuhe and Yang, Rui and Lie, Sui An and Lim, Taylor Xin Yi and Ning, Yilin and Li, Irene and Abdullah, Hairil Rizal and Ting, Daniel Shu Wei and Liu, Nan},
  journal={Journal of Medical Internet Research},
  volume={26},
  pages={e59439},
  year={2024},
  publisher={JMIR Publications Toronto, Canada}
}

@INPROCEEDINGS{11029967,
  author={Saghi, Zeinabsadat and Zimmermann, Thomas and Chattopadhyay, Souti},
  booktitle={2025 IEEE/ACM 47th International Conference on Software Engineering (ICSE)}, 
  title={Code Today, Deadline Tomorrow: Procrastination Among Software Developers}, 
  year={2025},
  volume={},
  number={},
  pages={1204-1216},
  doi={10.1109/ICSE55347.2025.00198}}

@article{eibl2025exploring,
  title={Exploring the Challenges and Opportunities of AI-assisted Codebase Generation},
  author={Eibl, Philipp and Sabouri, Sadra and Chattopadhyay, Souti},
  journal={arXiv preprint arXiv:2508.07966},
  year={2025}
}

@inproceedings{chattopadhyay2020tale,
  title={A tale from the trenches: cognitive biases and software development},
  author={Chattopadhyay, Souti and Nelson, Nicholas and Au, Audrey and Morales, Natalia and Sanchez, Christopher and Pandita, Rahul and Sarma, Anita},
  booktitle={Proceedings of the ACM/IEEE 42nd International Conference on Software Engineering},
  pages={654--665},
  year={2020}
}

@article{mohanani2018cognitive,
  title={Cognitive biases in software engineering: A systematic mapping study},
  author={Mohanani, Rahul and Salman, Iflaah and Turhan, Burak and Rodr{\'\i}guez, Pilar and Ralph, Paul},
  journal={IEEE Transactions on Software Engineering},
  volume={46},
  number={12},
  pages={1318--1339},
  year={2018},
  publisher={IEEE}
}

@article{rastogi2022deciding,
  title={Deciding fast and slow: The role of cognitive biases in AI-assisted decision-making},
  author={Rastogi, Charvi and Zhang, Yunfeng and Wei, Dennis and Varshney, Kush R and Dhurandhar, Amit and Tomsett, Richard},
  journal={Proceedings of the ACM on Human-computer Interaction},
  volume={6},
  number={CSCW1},
  pages={1--22},
  year={2022},
  publisher={ACM New York, NY, USA}
}

@article{Wang2019Designing,title={Designing Theory-Driven User-Centric Explainable AI},author={Danding Wang and Qian Yang and Ashraf Abdul and Brian Y. Lim},journal={Proceedings of the 2019 CHI Conference on Human Factors in Computing Systems},year={2019},doi={10.1145/3290605.3300831}}

@article{Fronza2022KeepingFA,
  title={Keeping Fun Alive: an Experience Report on Running Online Coding Camps},
  author={Ilenia Fronza and Luis Corral and Xiaofeng Wang and Claus Pahl},
  journal={2022 IEEE/ACM 44th International Conference on Software Engineering: Software Engineering Education and Training (ICSE-SEET)},
  year={2022},
  pages={165-175},
  url={https://api.semanticscholar.org/CorpusID:247187874}
}

@incollection{HANSKORTELING2022610,
title = {Cognitive Biases},
editor = {Sergio {Della Sala}},
booktitle = {Encyclopedia of Behavioral Neuroscience, 2nd edition (Second Edition)},
publisher = {Elsevier},
edition = {Second Edition},
address = {Oxford},
pages = {610-619},
year = {2022},
isbn = {978-0-12-821636-1},
doi = {https://doi.org/10.1016/B978-0-12-809324-5.24105-9},
url = {https://www.sciencedirect.com/science/article/pii/B9780128093245241059},
author = {J.E. {(Hans) Korteling} and Alexander Toet},
}

@article{furnham2011literature,
  title={A literature review of the anchoring effect},
  author={Furnham, Adrian and Boo, Hua Chu},
  journal={The journal of socio-economics},
  volume={40},
  number={1},
  pages={35--42},
  year={2011},
  publisher={Elsevier}
}

@article{tversky1974judgment,
  title={Judgment under Uncertainty: Heuristics and Biases: Biases in judgments reveal some heuristics of thinking under uncertainty.},
  author={Tversky, Amos and Kahneman, Daniel},
  journal={science},
  volume={185},
  number={4157},
  pages={1124--1131},
  year={1974},
  publisher={American association for the advancement of science}
}

@article{baron1988outcome,
  title={Outcome bias in decision evaluation.},
  author={Baron, Jonathan and Hershey, John C},
  journal={Journal of personality and social psychology},
  volume={54},
  number={4},
  pages={569},
  year={1988},
  publisher={American Psychological Association}
}

@inproceedings{Echterhoff2024CognitiveBI,
  title={Cognitive Bias in Decision-Making with LLMs},
  author={Jessica Maria Echterhoff and Yao Liu and Abeer Alessa and Julian McAuley and Zexue He},
  booktitle={Conference on Empirical Methods in Natural Language Processing},
  year={2024},
  url={https://api.semanticscholar.org/CorpusID:268230909}
}

@article{Gerlich2025AITI,
  title={AI Tools in Society: Impacts on Cognitive Offloading and the Future of Critical Thinking},
  author={Michael Gerlich},
  journal={Societies},
  year={2025},
  url={https://api.semanticscholar.org/CorpusID:275307388}
}

@article{sharot2007neural,
  title={Neural mechanisms mediating optimism bias},
  author={Sharot, Tali and Riccardi, Alison M and Raio, Candace M and Phelps, Elizabeth A},
  journal={Nature},
  volume={450},
  number={7166},
  pages={102--105},
  year={2007},
  publisher={Nature Publishing Group UK London}
}

@article{milgram1963behavioral,
  title={Behavioral study of obedience.},
  author={Milgram, Stanley},
  journal={The Journal of abnormal and social psychology},
  volume={67},
  number={4},
  pages={371},
  year={1963},
  publisher={American Psychological Association}
}

@article{tversky1981framing,
  title={The framing of decisions and the psychology of choice},
  author={Tversky, Amos and Kahneman, Daniel},
  journal={science},
  volume={211},
  number={4481},
  pages={453--458},
  year={1981},
  publisher={American Association for the Advancement of Science}
}

@article{thaler1981some,
  title={Some empirical evidence on dynamic inconsistency},
  author={Thaler, Richard},
  journal={Economics letters},
  volume={8},
  number={3},
  pages={201--207},
  year={1981},
  publisher={Elsevier}
}

@article{sunstein2002probability,
  title={Probability neglect: Emotions, worst cases, and law},
  author={Sunstein, Cass R},
  journal={Yale Lj},
  volume={112},
  pages={61},
  year={2002},
  publisher={HeinOnline}
}

@article{hsee2004distinction,
  title={Distinction bias: misprediction and mischoice due to joint evaluation.},
  author={Hsee, Christopher K and Zhang, Jiao},
  journal={Journal of personality and social psychology},
  volume={86},
  number={5},
  pages={680},
  year={2004},
  publisher={American Psychological Association}
}

@article{hsee1998less,
  title={Less is better: When low-value options are valued more highly than high-value options},
  author={Hsee, Christopher K},
  journal={Journal of Behavioral Decision Making},
  volume={11},
  number={2},
  pages={107--121},
  year={1998},
  publisher={Wiley Online Library}
}

@article{morewedge2015explanations,
  title={Explanations of the endowment effect: an integrative review},
  author={Morewedge, Carey K and Giblin, Colleen E},
  journal={Trends in cognitive sciences},
  volume={19},
  number={6},
  pages={339--348},
  year={2015},
  publisher={Elsevier}
}

@article{staw1976knee,
  title={Knee-deep in the big muddy: A study of escalating commitment to a chosen course of action},
  author={Staw, Barry M},
  journal={Organizational behavior and human performance},
  volume={16},
  number={1},
  pages={27--44},
  year={1976},
  publisher={Elsevier}
}

@article{german2005functional,
  title={Functional fixedness in a technologically sparse culture},
  author={German, Tim P and Barrett, H Clark},
  journal={Psychological science},
  volume={16},
  number={1},
  pages={1--5},
  year={2005},
  publisher={SAGE Publications Sage CA: Los Angeles, CA}
}

@article{zajonc2001mere,
  title={Mere exposure: A gateway to the subliminal},
  author={Zajonc, Robert B},
  journal={Current directions in psychological science},
  volume={10},
  number={6},
  pages={224--228},
  year={2001},
  publisher={SAGE Publications Sage CA: Los Angeles, CA}
}

@article{mortell2013physician,
  title={Physician ‘defiance’towards hand hygiene compliance: Is there a theory--practice--ethics gap?},
  author={Mortell, Manfred and Balkhy, Hanan H and Tannous, Elias B and Jong, Mei Thiee},
  journal={Journal of the Saudi Heart Association},
  volume={25},
  number={3},
  pages={203--208},
  year={2013},
  publisher={Elsevier}
}

@article{samuelson1988status,
  title={Status quo bias in decision making},
  author={Samuelson, William and Zeckhauser, Richard},
  journal={Journal of risk and uncertainty},
  volume={1},
  pages={7--59},
  year={1988},
  publisher={Springer}
}

@article{jackson1982empirical,
  title={An empirical study of travel time variability and travel choice behavior},
  author={Jackson, W Burke and Jucker, James V},
  journal={Transportation Science},
  volume={16},
  number={4},
  pages={460--475},
  year={1982},
  publisher={INFORMS}
}

@article{sharot2011optimism,
  title={The optimism bias},
  author={Sharot, Tali},
  journal={Current biology},
  volume={21},
  number={23},
  pages={R941--R945},
  year={2011},
  publisher={Elsevier}
}

@article{norton2012ikea,
  title={The IKEA effect: When labor leads to love},
  author={Norton, Michael I and Mochon, Daniel and Ariely, Dan},
  journal={Journal of consumer psychology},
  volume={22},
  number={3},
  pages={453--460},
  year={2012},
  publisher={Elsevier}
}

@article{ross1991barriers,
  title={Barriers to conflict resolution},
  author={Ross, Lee and Stillinger, Constance},
  journal={Negot. J.},
  volume={7},
  pages={389},
  year={1991},
  publisher={HeinOnline}
}

@article{kahneman1972subjective,
  title={Subjective probability: A judgment of representativeness},
  author={Kahneman, Daniel and Tversky, Amos},
  journal={Cognitive psychology},
  volume={3},
  number={3},
  pages={430--454},
  year={1972},
  publisher={Elsevier}
}

@article{mahoney1977publication,
  title={Publication prejudices: An experimental study of confirmatory bias in the peer review system},
  author={Mahoney, Michael J},
  journal={Cognitive therapy and research},
  volume={1},
  pages={161--175},
  year={1977},
  publisher={Springer}
}

@article{chapman1969illusory,
  title={Illusory correlation as an obstacle to the use of valid psychodiagnostic signs.},
  author={Chapman, Loren J and Chapman, Jean P},
  journal={Journal of abnormal psychology},
  volume={74},
  number={3},
  pages={271},
  year={1969},
  publisher={American Psychological Association}
}

@article{voss2012potato,
  title={The potato chip really does look like Elvis! Neural hallmarks of conceptual processing associated with finding novel shapes subjectively meaningful},
  author={Voss, Joel L and Federmeier, Kara D and Paller, Ken A},
  journal={Cerebral Cortex},
  volume={22},
  number={10},
  pages={2354--2364},
  year={2012},
  publisher={Oxford University Press}
}

@article{allison1985group,
  title={The group attribution error},
  author={Allison, Scott T and Messick, David M},
  journal={Journal of Experimental Social Psychology},
  volume={21},
  number={6},
  pages={563--579},
  year={1985},
  publisher={Elsevier}
}

@article{ross1979egocentric,
  title={Egocentric biases in availability and attribution.},
  author={Ross, Michael and Sicoly, Fiore},
  journal={Journal of personality and social psychology},
  volume={37},
  number={3},
  pages={322},
  year={1979},
  publisher={American Psychological Association}
}

@article{nisbett1977halo,
  title={THE HALO EFFECTS: EVIDENCE FOR UNCONSCIOUS ALTERATION OF JUDGMENTS.},
  author={Nisbett, RE and TD, WILSON},
  year={1977}
}

@article{correll2002police,
  title={The police officer's dilemma: using ethnicity to disambiguate potentially threatening individuals.},
  author={Correll, Joshua and Park, Bernadette and Judd, Charles M and Wittenbrink, Bernd},
  journal={Journal of personality and social psychology},
  volume={83},
  number={6},
  pages={1314},
  year={2002},
  publisher={American Psychological Association}
}

@article{waytz2010sees,
  title={Who sees human? The stability and importance of individual differences in anthropomorphism},
  author={Waytz, Adam and Cacioppo, John and Epley, Nicholas},
  journal={Perspectives on Psychological Science},
  volume={5},
  number={3},
  pages={219--232},
  year={2010},
  publisher={Sage Publications Sage CA: Los Angeles, CA}
}

@article{hedlund2000risky,
  title={Risky business: safety regulations, risk compensation, and individual behavior},
  author={Hedlund, James},
  journal={Injury prevention},
  volume={6},
  number={2},
  pages={82--89},
  year={2000},
  publisher={BMJ Publishing Group Ltd}
}

@article{sparrow2011google,
  title={Google effects on memory: Cognitive consequences of having information at our fingertips},
  author={Sparrow, Betsy and Liu, Jenny and Wegner, Daniel M},
  journal={science},
  volume={333},
  number={6043},
  pages={776--778},
  year={2011},
  publisher={American Association for the Advancement of Science}
}

@article{Felin2024TheoryIA,
  title={Theory Is All You Need: AI, Human Cognition, and Causal Reasoning},
  author={Teppo Felin and Matthias Holweg},
  journal={Strategy Science},
  year={2024},
  url={https://api.semanticscholar.org/CorpusID:274493956}
}

@article{taylor1981self,
  title={Self-serving and group-serving bias in attribution},
  author={Taylor, Donald M and Doria, Janet R},
  journal={The Journal of Social Psychology},
  volume={113},
  number={2},
  pages={201--211},
  year={1981},
  publisher={Taylor \& Francis}
}

@article{pennycook2018prior,
  title={Prior exposure increases perceived accuracy of fake news.},
  author={Pennycook, Gordon and Cannon, Tyrone D and Rand, David G},
  journal={Journal of experimental psychology: general},
  volume={147},
  number={12},
  pages={1865},
  year={2018},
  publisher={American Psychological Association}
}

@article{goddard2011automation,
  title={Automation bias--a hidden issue for clinical decision support system use},
  author={Goddard, Kate and Roudsari, Abdul and Wyatt, Jeremy C},
  journal={International perspectives in health informatics},
  pages={17--22},
  year={2011},
  publisher={IOS Press}
}

@article{moore1983archaeology,
  title={Archaeology and the law of the hammer},
  author={Moore, James A and Keene, Arthur S},
  journal={Archaeological Hammers and theories},
  pages={3--13},
  year={1983},
  publisher={Elsevier}
}

@article{oberai2018unconscious,
  title={Unconscious bias: thinking without thinking},
  author={Oberai, Himani and Anand, Ila Mehrotra},
  journal={Human resource management international digest},
  volume={26},
  number={6},
  pages={},
  year={2018},
  publisher={Emerald Publishing Limited}
}

@book{pohl2004cognitive,
  title={Cognitive illusions: A handbook on fallacies and biases in thinking, judgement and memory},
  author={Pohl, R{\"u}diger},
  year={2004},
  publisher={Psychology press}
}

@article{tuccio2011heuristics,
  title={Heuristics to improve human factors performance in aviation},
  author={Tuccio, William A},
  journal={Journal of Aviation/Aerospace Education \& Research},
  volume={20},
  number={3},
  pages={8},
  year={2011},
  publisher={Embry-Riddle Aeronautical University}
}

@article{kahneman2000evaluation,
  title={Evaluation by moments: Past and future},
  author={Kahneman, Daniel},
  journal={Choices, values, and frames},
  pages={693--708},
  year={2000},
  publisher={Citeseer}
}

@article{herrmann2011effect,
  title={The effect of default options on choice—Evidence from online product configurators},
  author={Herrmann, Andreas and Goldstein, Daniel G and Stadler, Rupert and Landwehr, Jan R and Heitmann, Mark and Hofstetter, Reto and Huber, Frank},
  journal={Journal of Retailing and Consumer Services},
  volume={18},
  number={6},
  pages={483--491},
  year={2011},
  publisher={Elsevier}
}

@article{forsyth2008allocating,
  title={Allocating time to future tasks: The effect of task segmentation on planning fallacy bias},
  author={Forsyth, Darryl K and Burt, Christopher DB},
  journal={Memory \& cognition},
  volume={36},
  number={4},
  pages={791--798},
  year={2008},
  publisher={Springer}
}

@article{shmueli2016can,
  title={Can the outside-view approach improve planning decisions in software development projects?},
  author={Shmueli, Ofira and Pliskin, Nava and Fink, Lior},
  journal={Information Systems Journal},
  volume={26},
  number={4},
  pages={395--418},
  year={2016},
  publisher={Wiley Online Library}
}

@article{liden2019devil,
  title={From devil’s advocate to crime fighter: Confirmation bias and debiasing techniques in prosecutorial decision-making},
  author={Lid{\'e}n, Moa and Gr{\"a}ns, Minna and Juslin, Peter},
  journal={Psychology, Crime \& Law},
  volume={25},
  number={5},
  pages={494--526},
  year={2019},
  publisher={Taylor \& Francis}
}

@article{Kinsey2020BurningBM,
  title={Burning biases: Mitigating cognitive biases in fire engineering},
  author={Michael J. Kinsey and Max Kinateder and Steve M. V. Gwynne and Danny Hopkin},
  journal={Fire and Materials},
  year={2020},
  volume={45},
  pages={543 - 552},
  url={https://api.semanticscholar.org/CorpusID:216339290}
}

@book{drabek2012human,
  title={Human system responses to disaster: An inventory of sociological findings},
  author={Drabek, Thomas E},
  year={2012},
  publisher={Springer Science \& Business Media}
}

@incollection{rogers2014diffusion,
  title={Diffusion of innovations},
  author={Rogers, Everett M and Singhal, Arvind and Quinlan, Margaret M},
  booktitle={An integrated approach to communication theory and research},
  pages={432--448},
  year={2014},
  publisher={Routledge}
}

@article{chakraborty2021present,
  title={Present bias},
  author={Chakraborty, Anujit},
  journal={Econometrica},
  volume={89},
  number={4},
  pages={1921--1961},
  year={2021},
  publisher={Wiley Online Library}
}

@article{ioannidis2005most,
  title={Why most published research findings are false},
  author={Ioannidis, John PA},
  journal={PLoS medicine},
  volume={2},
  number={8},
  pages={e124},
  year={2005},
  publisher={Public Library of Science}
}

@article{mike2022common,
  title={What is common to transportation and health in machine learning education? The domain neglect bias},
  author={Mike, Koby and Hazzan, Orit},
  journal={IEEE Transactions on Education},
  volume={66},
  number={3},
  pages={226--233},
  year={2022},
  publisher={IEEE}
}

@article{schenk2011exploiting,
  title={Exploiting the salience bias in designing taxes},
  author={Schenk, Deborah H},
  journal={Yale J. on Reg.},
  volume={28},
  pages={253},
  year={2011},
  publisher={HeinOnline}
}

@book{plous1993psychology,
  title={The psychology of judgment and decision making.},
  author={Plous, Scott},
  year={1993},
  publisher={Mcgraw-Hill Book Company}
}

@article{glenberg1980two,
  title={A two-process account of long-term serial position effects.},
  author={Glenberg, Arthur M and Bradley, Margaret M and Stevenson, Jennifer A and Kraus, Thomas A and Tkachuk, Marilyn J and Gretz, Ann L and Fish, Joel H and Turpin, BettyAnn M},
  journal={Journal of Experimental Psychology: Human Learning and Memory},
  volume={6},
  number={4},
  pages={355},
  year={1980},
  publisher={American Psychological Association}
}

@article{baddeley1993recency,
  title={The recency effect: Implicit learning with explicit retrieval?},
  author={Baddeley, Alan D and Hitch, Graham},
  journal={Memory \& cognition},
  volume={21},
  pages={146--155},
  year={1993},
  publisher={Springer}
}

@article{neighbors2019cognitive,
  title={Cognitive factors and addiction},
  author={Neighbors, Clayton and Tomkins, Mary M and Riggs, Jordanna Lembo and Angosta, Joanne and Weinstein, Andrew P},
  journal={Current opinion in psychology},
  volume={30},
  pages={128--133},
  year={2019},
  publisher={Elsevier}
}

@article{macleod2007concept,
  title={The concept of inhibition in cognition.},
  author={MacLeod, Colin M},
  year={2007},
  publisher={American Psychological Association}
}

@misc{conrad1958onset,
  title={The onset of schizophrenia: an attempt to form an analysis of delusion},
  author={Conrad, Klaus},
  year={1958},
  publisher={German. Georg Thieme Verlag}
}

@article{gray2010blaming,
  title={Blaming God for our pain: Human suffering and the divine mind},
  author={Gray, Kurt and Wegner, Daniel M},
  journal={Personality and Social Psychology Review},
  volume={14},
  number={1},
  pages={7--16},
  year={2010},
  publisher={Sage Publications Sage CA: Los Angeles, CA}
}

@article{coley2012common,
  title={Common origins of diverse misconceptions: Cognitive principles and the development of biology thinking},
  author={Coley, John D and Tanner, Kimberly D},
  journal={CBE—Life Sciences Education},
  volume={11},
  number={3},
  pages={209--215},
  year={2012},
  publisher={American Society for Cell Biology}
}

@article{sabouri2025trust,
  title={Trust dynamics in AI-assisted development: Definitions, factors, and implications},
  author={Sabouri, Sadra and Eibl, Philipp and Zhou, Xinyi and Ziyadi, Morteza and Medvidovic, Nenad and Lindemann, Lars and Chattopadhyay, Souti},
  year={2025}
}

@inproceedings{ross2023programmer,
  title={The programmer’s assistant: Conversational interaction with a large language model for software development},
  author={Ross, Steven I and Martinez, Fernando and Houde, Stephanie and Muller, Michael and Weisz, Justin D},
  booktitle={Proceedings of the 28th International Conference on Intelligent User Interfaces},
  pages={491--514},
  year={2023}
}

@inproceedings{sallou2024breaking,
  title={Breaking the silence: the threats of using llms in software engineering},
  author={Sallou, June and Durieux, Thomas and Panichella, Annibale},
  booktitle={Proceedings of the 2024 ACM/IEEE 44th International Conference on Software Engineering: New Ideas and Emerging Results},
  pages={102--106},
  year={2024}
}

@article{jin2024llms,
  title={From llms to llm-based agents for software engineering: A survey of current, challenges and future},
  author={Jin, Haolin and Huang, Linghan and Cai, Haipeng and Yan, Jun and Li, Bo and Chen, Huaming},
  journal={arXiv preprint arXiv:2408.02479},
  year={2024}
}

@article{nadeau1993new,
  title={New evidence about the existence of a bandwagon effect in the opinion formation process},
  author={Nadeau, Richard and Cloutier, Edouard and Guay, J-H},
  journal={International Political Science Review},
  volume={14},
  number={2},
  pages={203--213},
  year={1993},
  publisher={Sage Publications Sage CA: Thousand Oaks, CA}
}

@inproceedings{arteaga2024support,
  title={How to support ml end-user programmers through a conversational agent},
  author={Arteaga Garcia, Emily Judith and Nicolaci Pimentel, Jo{\~a}o Felipe and Feng, Zixuan and Gerosa, Marco and Steinmacher, Igor and Sarma, Anita},
  booktitle={Proceedings of the 46th IEEE/ACM International Conference on Software Engineering},
  pages={1--12},
  year={2024}
}

@article{nagappan2008realizing,
  title={Realizing quality improvement through test driven development: results and experiences of four industrial teams},
  author={Nagappan, Nachiappan and Maximilien, E Michael and Bhat, Thirumalesh and Williams, Laurie},
  journal={Empirical Software Engineering},
  volume={13},
  pages={289--302},
  year={2008},
  publisher={Springer}
}

@article{barke2023grounded,
  title={Grounded copilot: How programmers interact with code-generating models},
  author={Barke, Shraddha and James, Michael B and Polikarpova, Nadia},
  journal={Proceedings of the ACM on Programming Languages},
  volume={7},
  number={OOPSLA1},
  pages={85--111},
  year={2023},
  publisher={ACM New York, NY, USA}
}

@article{ha2024improving,
  title={Improving trust in AI with mitigating confirmation bias: Effects of explanation type and debiasing strategy for decision-making with explainable AI},
  author={Ha, Taehyun and Kim, Sangyeon},
  journal={International journal of human--computer interaction},
  volume={40},
  number={24},
  pages={8562--8573},
  year={2024},
  publisher={Taylor \& Francis}
}

@book{baron2023thinking,
  title={Thinking and deciding},
  author={Baron, Jonathan},
  year={2023},
  publisher={Cambridge University Press}
}

@article{mathew1994qualitative,
  title={Qualitative data analysis},
  author={Mathew, B Miles and Huberman, A Michael},
  journal={An expanded sourcebook (2nd edition) Sage Publication},
  year={1994}
}

@inproceedings{meyer2014software,
  title={Software developers' perceptions of productivity},
  author={Meyer, Andr{\'e} N and Fritz, Thomas and Murphy, Gail C and Zimmermann, Thomas},
  booktitle={Proceedings of the 22nd ACM SIGSOFT International Symposium on Foundations of Software Engineering},
  pages={19--29},
  year={2014}
}

@inproceedings{minelli2015know,
  title={I know what you did last summer-an investigation of how developers spend their time},
  author={Minelli, Roberto and Mocci, Andrea and Lanza, Michele},
  booktitle={2015 IEEE 23rd international conference on program comprehension},
  pages={25--35},
  year={2015},
  organization={IEEE}
}

@inproceedings{calikli2010empirical,
author = {Calikli, Gul and Bener, Ayse},
title = {Empirical analyses of the factors affecting confirmation bias and the effects of confirmation bias on software developer/tester performance},
year = {2010},
isbn = {9781450304047},
publisher = {Association for Computing Machinery},
address = {New York, NY, USA},
url = {https://doi.org/10.1145/1868328.1868344},
doi = {10.1145/1868328.1868344},
booktitle = {Proceedings of the 6th International Conference on Predictive Models in Software Engineering},
articleno = {10},
numpages = {11},
location = {Timi\c{s}oara, Romania},
series = {PROMISE '10}
}

@article{buehler1994exploring,
  title={Exploring the" planning fallacy": Why people underestimate their task completion times.},
  author={Buehler, Roger and Griffin, Dale and Ross, Michael},
  journal={Journal of personality and social psychology},
  volume={67},
  number={3},
  pages={366},
  year={1994},
  publisher={American Psychological Association}
}

@article{millar1997effects,
  title={The effects of cognitive capacity and suspicion on truth bias},
  author={Millar, Murray G and Millar, Karen U},
  journal={Communication Research},
  volume={24},
  number={5},
  pages={556--570},
  year={1997},
  publisher={Sage Publications London}
}

@article{frisch1988ambiguity,
  title={Ambiguity and rationality},
  author={Frisch, Deborah and Baron, Jonathan},
  journal={Journal of Behavioral Decision Making},
  volume={1},
  number={3},
  pages={149--157},
  year={1988},
  publisher={Wiley Online Library}
}

@inproceedings{nam2024usinganllm,
author = {Nam, Daye and Macvean, Andrew and Hellendoorn, Vincent and Vasilescu, Bogdan and Myers, Brad},
title = {Using an LLM to Help With Code Understanding},
year = {2024},
isbn = {9798400702174},
publisher = {Association for Computing Machinery},
address = {New York, NY, USA},
url = {https://doi.org/10.1145/3597503.3639187},
doi = {10.1145/3597503.3639187},
booktitle = {Proceedings of the IEEE/ACM 46th International Conference on Software Engineering},
articleno = {97},
numpages = {13},
location = {Lisbon, Portugal},
series = {ICSE '24}
}

@inproceedings{lee2025impact,
  title={The impact of generative AI on critical thinking: Self-reported reductions in cognitive effort and confidence effects from a survey of knowledge workers},
  author={Lee, Hao-Ping and Sarkar, Advait and Tankelevitch, Lev and Drosos, Ian and Rintel, Sean and Banks, Richard and Wilson, Nicholas},
  booktitle={Proceedings of the 2025 CHI conference on human factors in computing systems},
  pages={1--22},
  year={2025}
}

@inproceedings{subramonyam2024bridging,
  title={Bridging the gulf of envisioning: Cognitive challenges in prompt based interactions with llms},
  author={Subramonyam, Hari and Pea, Roy and Pondoc, Christopher and Agrawala, Maneesh and Seifert, Colleen},
  booktitle={Proceedings of the 2024 CHI Conference on Human Factors in Computing Systems},
  pages={1--19},
  year={2024}
}

@article{su2024dualformer,
  title={Dualformer: Controllable fast and slow thinking by learning with randomized reasoning traces},
  author={Su, DiJia and Sukhbaatar, Sainbayar and Rabbat, Michael and Tian, Yuandong and Zheng, Qinqing},
  journal={arXiv preprint arXiv:2410.09918},
  year={2024}
}

@inproceedings{pareek2024trustdev,
author = {Pareek, Saumya and Velloso, Eduardo and Goncalves, Jorge},
title = {Trust Development and Repair in AI-Assisted Decision-Making during Complementary Expertise},
year = {2024},
isbn = {9798400704505},
publisher = {Association for Computing Machinery},
address = {New York, NY, USA},
url = {https://doi-org.libproxy2.usc.edu/10.1145/3630106.3658924},
doi = {10.1145/3630106.3658924},
booktitle = {Proceedings of the 2024 ACM Conference on Fairness, Accountability, and Transparency},
pages = {546–561},
numpages = {16},
location = {Rio de Janeiro, Brazil},
series = {FAccT '24}
}

@article{fasolo2025mitigating,
  title={Mitigating cognitive bias to improve organizational decisions: An integrative review, framework, and research agenda},
  author={Fasolo, Barbara and Heard, Claire and Scopelliti, Irene},
  journal={Journal of Management},
  volume={51},
  number={6},
  pages={2182--2211},
  year={2025},
  publisher={SAGE Publications Sage CA: Los Angeles, CA}
}

@article{Shah2024A,title={A Comprehensive Review of Bias in Deep Learning Models: Methods, Impacts, and Future Directions},author={Milind Shah and Nitesh M. Sureja},journal={Archives of Computational Methods in Engineering},year={2024},doi={10.1007/s11831-024-10134-2}}

@article{Gray2023Measurement,title={Measurement and Mitigation of Bias in Artificial Intelligence: A Narrative Literature Review for Regulatory Science},author={Magnus Gray and Ravi K Samala and Qi Liu and Denny Skiles and Joshua Xu and Weida Tong and Leihong Wu},journal={Clinical Pharmacology \& Therapeutics},year={2023},volume={115},doi={10.1002/cpt.3117}}

@article{Hasanzadeh2025BiasRA,
  title={Bias recognition and mitigation strategies in artificial intelligence healthcare applications},
  author={Fereshteh Hasanzadeh and Colin B Josephson and Gabriella Waters and Demilade A. Adedinsewo and Zahra Azizi and James A. White},
  journal={NPJ Digital Medicine},
  year={2025},
  volume={8},
  url={https://api.semanticscholar.org/CorpusID:276932763}
}

@article{inan2025better,
  title={Better slow than sorry: Introducing positive friction for reliable dialogue systems},
  author={Inan, Mert and Sicilia, Anthony and Dey, Suvodip and Dongre, Vardhan and Srinivasan, Tejas and Thomason, Jesse and T{\"u}r, G{\"o}khan and Hakkani-T{\"u}r, Dilek and Alikhani, Malihe},
  journal={arXiv preprint arXiv:2501.17348},
  year={2025}
}

@inproceedings{echterhoff2022ai,
  title={AI-moderated decision-making: Capturing and balancing anchoring bias in sequential decision tasks},
  author={Echterhoff, Jessica Maria and Yarmand, Matin and McAuley, Julian},
  booktitle={Proceedings of the 2022 CHI Conference on Human Factors in Computing Systems},
  pages={1--9},
  year={2022}
}

@inproceedings{liu2024ai,
  title={How ai processing delays foster creativity: Exploring research question co-creation with an llm-based agent},
  author={Liu, Yiren and Chen, Si and Cheng, Haocong and Yu, Mengxia and Ran, Xiao and Mo, Andrew and Tang, Yiliu and Huang, Yun},
  booktitle={Proceedings of the 2024 CHI Conference on Human Factors in Computing Systems},
  pages={1--25},
  year={2024}
}

@inproceedings{cheng2019explaining,
  title={Explaining decision-making algorithms through UI: Strategies to help non-expert stakeholders},
  author={Cheng, Hao-Fei and Wang, Ruotong and Zhang, Zheng and O'connell, Fiona and Gray, Terrance and Harper, F Maxwell and Zhu, Haiyi},
  booktitle={Proceedings of the 2019 chi conference on human factors in computing systems},
  pages={1--12},
  year={2019}
}

\end{document}